\documentclass[12pt]{article}
\usepackage{amsmath}
\usepackage[dvips]{graphicx}
\usepackage{epsfig}
\usepackage{amsmath}
\usepackage{amssymb}
\usepackage{float}
\usepackage{graphics,epstopdf}
\usepackage{amsfonts}
\usepackage{amsthm}
\usepackage[usenames]{color}

\definecolor{AV}{rgb}{0.65,0.0,0}
\definecolor{GC}{rgb}{0,0.0,0.65}
\definecolor{WS}{rgb}{0,0.65,0}

\usepackage[
      colorlinks=true,
      linkcolor=blue,
      urlcolor=blue,
      filecolor=blue,
      citecolor=red,
      pdfstartview=FitV,
      pdftitle={},
      pdfauthor={},
      pdfsubject={},
      pdfkeywords={},
      pdfpagemode=None,
      bookmarksopen=true
]{hyperref}

\newcommand{\bm}{\begin{multiline}}
\newcommand{\beq}{\begin{equation}}
\newcommand{\eeq}{\end{equation}}
\newcommand{\beqs}{\begin{eqnarray}}
\newcommand{\eeqs}{\end{eqnarray}}

\newcommand{\ra}{\rightarrow}

\begin{document}

\begin{center}

\textbf{\Large Charged Particles Orbiting a Magnetized Black Hole Immersed in Quintessential Matter }

\vspace{48pt}

\textbf{Vitalie Lungu, }\footnote{Corresponding author e-mail: \texttt{vitalie.lungu@student.uaic.ro}}
\textbf{Marina-Aura Dariescu,}\footnote{E-mail: \texttt{marina@uaic.ro}}
\textbf{ Cristian Stelea,}\footnote{E-mail: \texttt{cristian.stelea@uaic.ro}}

\vspace*{0.2cm}

\textit{$^{1,2}$ Faculty of Physics, ``Alexandru Ioan Cuza" University of Iasi}\\[0pt]
\textit{11 Bd. Carol I, Iasi, 700506, Romania}\\[.5em]

\textit{$^3$ Department of Exact and Natural Sciences, Institute of Interdisciplinary Research,}\\[0pt]
\textit{``Alexandru Ioan Cuza" University of Iasi}\\[0pt]
\textit{11 Bd. Carol I, Iasi, 700506, Romania}\\[.5em]

\end{center}

\vspace{30pt}

\begin{abstract}
A new exact magnetized solution describing a Kiselev black hole immersed in a magnetic field is used for studying the dynamics of charged particles. Different types of trajectories are discussed. In the particular case of a weak magnetic field, we employ a first-order perturbative approach to analyze the perturbed circular orbits near the
minimum of the effective potential. We obtain an approximate solution for the bounded equatorial trajectory subjected to small radial and latitudinal oscillations.
The shape of the trajectory localized near the stable
circular orbit depends on the relation between the attractive gravitational force and the repulsive quintessence contribution.
\end{abstract}

\baselineskip 1.5em

\newpage

\section{Introduction}

Radio observations of synchrotron emission and polarization provide valuable information on the structure of magnetic fields in the interstellar medium of both nearby and distant galaxies \cite{Planck}. In case of black holes, there is a strong evidence for the presence of the electromagnetic field
around them \cite{Eat}. Even though, according to the ”no-hair theorem”, a black hole does not posses a magnetic field on its own, this
can be generated by the moving charged particles in the accretion disk, or by the surrounded
rotating matter \cite{Wald}. The dynamo mechanism from the accretion disk is responsible for the
creation of the magnetic field \cite{Frolov}.
Moreover, the magnetic field plays an important role in the
transfer of the radiation from the accretion disk to the jets \cite{Frolov, McK}.
On the other hand, it has
been found that the super massive black hole at the center of Milky Way, is surrounded by a
strong magnetic field which is not correlated to the accretion disk \cite{Eat}.

The estimated value of the magnetic induction may be $B_0 \sim 10^4$ (G) for supermassive black holes \cite{Daly}
and can reach $10^8$ (G) for stellar mass black holes \cite{Piotr}.
Even though these values cannot change the
spacetime geometry around the black holes, the corresponding Lorentz force acting on charged particles
has a strong influence on their trajectories. In this respect, the case of an axisymmetric and uniform at the spatial infinity magnetic field has been analyzed in
\cite{Frolov}.
In the more complex situation of an external combined magnetic field,
the equatorial and off-equatorial orbital motion of charged particles has been numerically investigated \cite{Ken}. It has been pointed out that the dynamics
of charged particles can be chaotic if the Lorenz force is comparable to the gravitational attraction of the black hole.

In what it concerns the spacetime in the vicinity of a black hole, the Ernst solution which describes a static Schwarzschild black hole immersed in an uniform magnetic field is the most popular one \cite{Ernst}. However, with the compelling evidence at all astrophysical scales for the existence of dark matter in galaxies \cite{Clowe}, more complex solutions have been derived by taking into account the dark sector contribution. The observations of the closest S-star to the Galaxy center have reported that the upper limit on the total dark mass inside the orbit is less than
$1 \%$ of the Sgr A* \cite{Grav}.

 Even though several scenarios have been proposed, the true nature of dark matter remains unclear. In case of quintessence \cite{Cap, Vik}, the solution obtained by Kiselev
\cite{Kis}
was extensively used for analyzing the null and timelike geodesics structure around the Schwarzschild black hole surrounded by quintessential matter \cite{Fer, Bad, Dar24}.

As it is known, the Kiselev geometry is sourced by an anisotropic fluid which satisfies the Equation of State $P^0 =w\rho^0$, where $P^0$ is the isotropic pressure $P^0 = (p_r^0 + p_{\theta}^0 + p_{\varphi}^0)/3$ and $\rho^0$ is the energy density. The equation of state parameter $w$ is in the range $w \in [ -1 , -1/3]$ in order to cause acceleration.

For $w=-2/3$, the Schwarzschild metric function $1-\frac{2M}{r}$ gets an additional linear contribution, $- kr$, where $k$ is the quintessence parameter. Recently, the Kiselev solution was reinterpreted as an exact solution of the so-called power-Maxwell theories \cite{Dar22}.

As a remarkable result, in 2022, Cardoso et al. found an exact solution that may serve as a model for a supermassive black hole in the center of a galaxy surrounded by a dark matter halo \cite{Cardoso}. The charged version of the Cardoso et al. solution was derived in \cite{SteleaPLB}, using the solution generating technique developed in \cite{SteleaPRD}. More recently, this solution generating technique has been generalized for metrics with axial symmetry in \cite{Stelea:2018elx}. 
In this work we shall make use of the above technique to add a poloidal magnetic field to a fluid solution using the results from \cite{Stelea:2018cgm}, \cite{Stelea:2018elx}, \cite{Yaz}. For this purpose, in the next section we give more details on the solution generating technique described in these papers and show how to generate a new solution: the magnetized Kiselev solution in four dimensions. This solution will allow us to achieve our main goal of this work, which is to analyze the motion of charged particles around a black
hole immersed in a magnetic field, in the presence of quintessence. 

The structure of the paper is as follows: After the Introduction, in section 2, we are describing the solution generating technique that will allow us to generate magnetized fluid solutions starting with spherically symmetric solutions of the Einstein-fluid equations, sourced by anisotropic fluids. In particular, in subsection $2.1$ we show how to generate the magnetized Kiselev black hole solution that will be used in the next sections. Throughout the paper, we consider the special case with the equation of state parameter $w = -2/3$, which leads to a simplicity of treatment.

In section 3, we are deriving the field equations for charged particles and obtain the corresponding trajectories using numerical methods.
A special attention is given to the effective potential which allows us to discuss and classify the particle's trajectories both in three dimensions and in the equatorial plane. It turns out that the shape of the potential and the type of trajectory are strongly influenced by the model's parameters $B_0$ and $k$.
In order to underline the influence of the quintessential matter on the test particle’s motion, we make a comparison with Melvin and Ernst magnetic universes. 

In section 4, we turn our attention to an analytical approach. In this respect, we use a simplified form of the field equations by assuming that the magnetic field is too weak to have a gravitational backreaction on the spacetime metric. However, the Lorentz force has a significant influence on the charged particle's trajectory \cite{Frolov} whose shape depends on the parameters $M$, $k$ and $B_0$. Firstly, the perturbed circular orbits in the equatorial plane are discussed in detail.
As expected, the obtained
trajectories show a curly trochoid-like behavior \cite{lim}. 
Secondly, within a three-dimensional approach, we analyze the quasi-harmonic particle's motion about the equatorial plane.
Using a perturbative approach, an approximate solution for the bounded equatorial trajectory subjected to small radial and latitudinal oscillations is obtained.
Since, in the $\theta-$direction, the motion is described by a Mathieu-type equation, we use the stability charts to determine the stability regions. In the vicinity of
the black hole, the gravitational field is very strong and together with the Lorentz force acts
on the test particle, whose motion may become chaotic. The final section is devoted to conclusions. 

Such theoretical studies on the dynamics of test particle can be compared to the available data on individual S2 stars orbiting around the massive and compact radio source located at the center of our
Galaxy (see for example \cite{Grav}). 

Throughout the paper, we work in the natural units for which $G=c=1$ and with the $+2$ signature for the spacetime metric.

\section{The magnetized Kiselev solution}

In this section we shall show how one can derive the magnetized solution starting from a seed solution sourced by an anisotropic fluid.

\subsection{The solution generating technique}
To understand the effects of the magnetizing technique on a fluid solution, let us first recall how it acts on a vacuum solution of the Einstein field equations.

Consider then a vacuum geometry that is a solution of the Einstein field equations of the form:
\beqs
ds^2&=&e^{2u}d\varphi^2+e^{-2u}H_{ij}dx^idx^j,
\label{3dvac}
\eeqs
where the function $u$ and, in general, the metric $H^{ij}$ could depend only on the remaining coordinates $x^i$ with $i=0..2$. However, since we are interested in static geometries we shall consider that the metric $H_{ij}$ depends only on the remaining coordinates $r$ and $\theta$. Einstein's field equations in four dimensions $R_{\mu\nu}=0$ can then be shown to be equivalent to the dimensionally reduced system of equations along the $\varphi$ direction:
\beqs
R_{ij}&=&2\nabla_iu\nabla_ju,~~~~~\nabla_i\nabla^iu=0,
\label{3dvacuum}
\eeqs
where $R_{ij}$ is the Ricci tensor of the three-dimensional geometry $H_{ij}$.

Suppose now that we add a magnetic field using the ansatz $A=\Phi d\varphi$. Then, the four dimensional solution 
\beqs
ds^2&=&e^{2U}d\varphi^2+e^{-2U}H_{ij}dx^idx^j,
\label{mag4d}
\eeqs
of the Einstein field equations $R_{\mu\nu}=8\pi T_{\mu\nu}^{em}$, with 
\beqs
T_{\mu\nu}^{em}&=&\frac{1}{4\pi}\left(F_{\mu\rho}F_{\nu}^{~\rho}-\frac{1}{4}g_{\mu\nu}F^2\right)
\label{Temfield}
\eeqs
being the stress-energy tensor for the electromagnetic field is also a solution of the effective field equations obtained by the dimensional reduction along the $\varphi$ direction:
\beqs
R_{ij}=2\nabla_iU\nabla_jU+2e^{-2U}\nabla_i\Phi\nabla_j\Phi,\nonumber\\
\nabla_i\left(e^{-2U}\nabla^i\Phi\right)=0,~~~~~\nabla_i(\nabla^i U)=-e^{-2U}\nabla_i\Phi\nabla^i\Phi.
\label{3dem}
\eeqs
where again $R_{ij}$ is the Ricci tensor of the three-dimensional geometry $H_{ij}$. Assuming now a functional dependence between the $U(\chi)$ and $\Phi(\chi)$ the last two equations in (\ref{3dem}) can be easily solved as (see also \cite{Chng:2006gh})
\beqs
e^{2U}&=&\frac{e^{2\chi}}{\left(1+B_0^2e^{2\chi}\right)^2},~~~~~\Phi=\frac{B_0e^{2\chi}}{1+B_0^2e^{2\chi}},
\label{emfields3d}
\eeqs
if $\nabla_i(\nabla^i\chi)=0$ and $B_0$ is a constant. Moreover, the first equation in (\ref{3dem}) can be recast into the form $R_{ij}=2\nabla_i\chi\nabla_j\chi$ and the system of equations (\ref{3dem}) is reduced to the system of equations (\ref{3dvacuum}) if one simply takes $\chi=u$ and assumes the functional dependence given in (\ref{emfields3d}). To conclude, from any vacuum solution of the Einstein field equations (\ref{3dvac}) one can easily construct the magnetized metric:
\beqs
ds^2&=&\frac{e^{2u}}{\Lambda^2}d\varphi^2+\Lambda^2e^{-2u}H_{ij}dx^idx^j
\eeqs
where we defined $\Lambda=1+B_0^2e^{2u}$, while the magnetic potential is simply given by $A_{\varphi}=\Phi=\frac{B_0e^{2u}}{\Lambda}$.

Now, let us take into account the presence of an anistropic fluid instead of a vacuum solution of the Einstein equations. The initial geometry can be again written in the form (\ref{3dvacuum}), however, let us assume at this point that it is sourced by an anistropic fluid described by the stress-energy tensor:
\beqs
T_{\mu\nu}^0&=&\rho^0u_{\mu}^0u_{\nu}^0+p_r^0\chi_{\mu}^0\chi_{\nu}^0+p_{t}^0(\xi_{\mu}^0\xi_{\nu}^0+\zeta_{\mu}^0\zeta_{\nu}^0),
\eeqs
where $u_{\mu}^0$ is the four-velocity of the fluid, $\chi_{\mu}^0$ is the unit vector in the radial direction, while $\xi_{\mu}^0$ and $\zeta_{\mu}^0$ are the corresponding unit vectors in the transverse directions $\theta$ and $\varphi$ respectively. Note that these unit vectors are defined using the initial geometry (\ref{3dvacuum}). Here $\rho^0$ is the energy density, $p_r^0$ is the pressure in the radial direction, while $p_{\theta}^0=p_{\varphi}^0=p_t^0$ is the anisotropic pressure in the transverse directions. Then Einstein's field equations in this case can be rewritten in the form:
\beqs
R_{\mu\nu}&=&8\pi\left(T_{\mu\nu}^0-\frac{1}{2}g_{\mu\nu}T^0\right),
\eeqs
where $T^0=-\rho^0+p_r^0+2p_t^0$ is the trace of the fluid stress-energy tensor.

If one performs now a dimensional reduction along the $\varphi$ direction one obtains the following field equations for the unmagnetized solution:
\beqs
R_{ij}&=&8\pi(p_r^0+p_t^0)e^{-2u}u_i^0u_j^0+8\pi(\rho^0-p_t^0)e^{-2u}\chi_i^0\chi_j^0+8\pi(\rho^0-p_r^0)e^{-2u}\xi_i^0\xi_j^0+2\nabla_iu\nabla_ju,\nonumber\\
\nabla_i(\nabla^iu)&=&4\pi(\rho^0-p_r^0)e^{-2u},
\label{nemagf}
\eeqs
where $R_{ij}$ is the Ricci tensor of the three-geometry $H_{ij}dx^idx^j$. Here we denoted by $u_i^0$ the four-velocity of the fluid, $\chi_i^0$ is the unit vector in the radial direction, while $\xi_i^0$ is the unit vector in the $\theta$ direction, all of them being computed now in the effective geometry $H_{ij}dx^idx^j$.

Consider now a magnetized solution of the Einstein-Maxwell-fluid field equations using the metric ansatz (\ref{mag4d}), while the ansatz for the electromagnetic field is again given by $A=\Phi d\varphi$. The Einstein-Maxwell-fluid equations can be again recast into the form:
\beqs
R_{\mu\nu}&=&8\pi\left(T_{\mu\nu}-\frac{1}{2}g_{\mu\nu}T\right)+8\pi T_{\mu\nu}^{em},
\eeqs
where now 
\beqs
T_{\mu\nu}&=&\rho u_{\mu}u_{\nu}+p_r\chi_{\mu}\chi_{\nu}+p_{t}(\xi_{\mu}\xi_{\nu}+\zeta_{\mu}\zeta_{\nu}),
\eeqs
is the corresponding fluid stress-energy tensor of the magnetized solution.

Performing now a dimensional reduction along the $\varphi$ direction one is lead to the effective system of equations:
\beqs
R_{ij}&=&8\pi(p_r+p_t)e^{-2U}u_i^0u_j^0+8\pi(\rho-p_t)e^{-2U}\chi_i^0\chi_j^0+4\pi(\rho-p_r)e^{-2U}\xi_i^0\xi_j^0\nonumber\\
&+&2\nabla_iU\nabla_jU+2e^{-2U}\nabla_i\Phi\nabla_j\Phi,\nonumber\\
\nabla_i(\nabla^iU)&=&4\pi(\rho-p_r)e^{-2U}-4\pi\sigma e^{-2U}-e^{-2U}\nabla_i\Phi\nabla^i\Phi,\nonumber\\
\nabla_i\left(e^{-2U}\nabla^i\Phi\right)&=&4\pi e^{-4U}J_{\varphi}.
\label{magnetfin}
\eeqs

If one assumes again a functional dependence between the magnetic potential $\Phi=\Phi(u)$ and the metric potential $U=U(u)$ of the form (\ref{emfields3d}), one finds that $2\nabla_iU\nabla_jU+2e^{-2U}\nabla_i\Phi\nabla_j\Phi=2\nabla_iu\nabla_ju$ such that the first equation in (\ref{magnetfin}) reduces to the first equation in (\ref{nemagf}) if one simply considers:
\beqs
\rho e^{-2U}&=&\rho^0e^{-2u},~~~p_re^{-2U}=p_r^0e^{-2u}, ~~~p_{t}e^{-2U}=p_t^0e^{-2u}.
\eeqs
Recall that from (\ref{emfields3d}) one has $e^{2U}=\frac{e^{2u}}{\Lambda^2}$ such that in the end:
\beqs
\rho&=&\frac{\rho^0}{\Lambda^2},~~~p_r=\frac{p_r^0}{\Lambda^2}, ~~~p_{t}=\frac{p_t^0}{\Lambda^2}.
\eeqs
The remaining two equations in (\ref{magnetfin}) will lead to the following expressions for $\sigma$ and $J_{\varphi}$:
\beqs
\sigma&=&-2(\rho-p_r)\frac{B_0^2e^{2u}}{\Lambda},\nonumber\\
J_{\varphi}&=&2(\rho-p_r)\frac{B_0e^{2u}}{\Lambda^2},
\eeqs
where we again defined $\Lambda=1+B_0^2e^{2u}$. Note that $\sigma\sim J_{\varphi}A^{\varphi}$ and it actually corresponds to the interaction term of the form $A_{\mu}J^{\mu}$ in the Lagrangean that leads to the Maxwell equation $F^{\mu\nu}_{~~;\nu}=4\pi J_{\mu}$. Since its contribution in the effective stress-energy tensor has the form $\sim J_{\mu}A_{\nu}\sim\sigma \zeta_{\mu}\zeta_{\nu}$ we have chosen here to include its contribution in the final fluid stress-energy tensor, such that the transverse pressure $p_{\varphi}$ is now modified to read $p_{\varphi}=\frac{p_t^0}{\Lambda^2}+\sigma$.

Finally, the magnetized geometry can be written as:
\beqs
ds^2&=&\frac{e^{2u}}{\Lambda^2}d\varphi^2+\Lambda^2e^{-2u}H_{ij}dx^idx^j.
\eeqs

\subsection{The magnetized Kiselev black hole solution}

The Kiselev geometry is described by the following static four-dimensional line-element  \cite{Kis}:
\beqs
ds^2&=&g_{\mu\nu}^0dx^{\mu}dx^{\nu}= - f(r) dt^2+\frac{dr^2}{f(r)} + r^2 ( d \theta^2 + \sin^2 \theta d \varphi^2)
\label{kiselev} \; ,
\eeqs
where
\begin{equation}
f(r)= 1 -\frac{2M}{r} - \frac{k}{r^{3 w +1}}.
\label{g0}
\end{equation}
Here $w$ is the equation of state parameter and $k$ is a positive quintessence parameter, which is related to the fluid quintessence energy density:
\beqs
\rho^0 &=& - \frac{3kw}{8\pi r^{3(w+1)}},
\eeqs
while the components of the anisotropic pressures can be written as $p_r^0=-\rho^0$ and the tangential pressures $p_t^0$ are given by:
\beqs
p_t^0&\equiv&p_{\theta}^0=p_{\varphi}^0=-\frac{3(3w+1)kw}{16 \pi r^{3(w+1)}}.
\eeqs
In order to have an accelerated expansion, the equation of state parameter $w$ should belong to the interval $w \in [ -1 , -1/3]$. 

Note now that the Kiselev solution is sourced by an anisotropic fluid \cite{Visser:2019brz}.  Indeed, it describes a  spherically symmetric solution of the Einstein-anisotropic fluid equations:
\beqs 
G_{\mu\nu}&=&8\pi T_{\mu\nu}^0,
\eeqs 
where the stress-energy $T_{\mu\nu}^0$ of the anisotropic distribution of matter has the form:
\beqs
T_{\mu\nu}^0&=&(\rho^0+p_t^0)u_{\mu}^0u_{\nu}^0+p_t^0g_{\mu\nu}^0+(p_r^0-p_t^0)\chi_{\mu}^0\chi_{\nu}^0.
\eeqs
Here $\rho^0$ is the fluid density, $p_r^0$ is the radial component of the pressure, while $p_t^0$ represents the transverse components of the pressure. Moreover, $u_{\mu}^0=-\sqrt{f(r)}\delta^t_{\mu}$ is the $4$-velocity of the fluid, while $(\chi^0)^{\mu}=\sqrt{f(r)}\delta^{\mu}_r$ is the unit spacelike vector in the radial direction.

To apply the magnetized technique to the Kiselev solution one should note that in this case:
\beqs
e^{2u}&=&r^2\sin^2\theta, ~~~e^{-2u}H_{ij}dx^idx^j=-f(r)dt^2+\frac{dr^2}{f(r)}+r^2d\theta^2.
\eeqs

As such, one could use the solution generating techniques developed in the previous subsection to obtain the following new exact magnetized solution described by the new geometry:
\begin{equation}
ds^2 = \Lambda^2\bigg[-f(r)dt^2+\frac{dr^2}{f(r)}+r^2d\theta^2\bigg]+\frac{r^2\sin^2\theta}{\Lambda^2}d\varphi^2,
\label{ds2}
\end{equation}
where
\begin{equation}
\Lambda = 1+B_0^2 r^2 \sin^2 \theta
\label{LambdaB}
\end{equation}
In the above relation, $B_0$ is a constant that can be related to the value of the poloidal magnetic field on the axis of symmetry. This axially-symmetric geometry is now sourced by an anisotropic fluid with the density and pressure components of the form:
\beqs
\rho&=&\frac{\rho^0}{\Lambda^2}, \quad p_r=\frac{p_{r}^0}{\Lambda^2}, \quad p_{\theta}=\frac{p_{\theta}^0}{\Lambda^2},\quad p_{\varphi}=\frac{p_{\varphi}^0}{\Lambda^2}+\sigma,
\eeqs
and also by an electromagnetic field sourced by the above fluid (as one can see from the corresponding Maxwell equations). More precisely, the poloidal magnetic field is generated by the following electromagnetic potential:
\begin{equation}
A_{\varphi} \, = \frac{B_0 r^2 \sin^2 \theta}{\Lambda}. 
\label{A3}
\end{equation}

Then the system of Einstein-Maxwell-fluid equations of motion:
\beqs
G_{\mu\nu}&=&8\pi T_{\mu\nu}^{fluid}+8\pi T_{\mu\nu}^{em},\nonumber\\
F^{\mu\nu}_{~~;\nu}&=&4\pi J^{\mu},~~~~J^{\mu}_{~;\mu}=0
\label{eomEMF}
\eeqs
is satisfied if one considers that the stress-energy tensor of the anisotropic fluid is given by:
\beqs
T_{\mu\nu}^{fluid}&=&\rho u_{\mu}u_{\nu}+p_r\chi_{\mu}\chi_{\nu}+p_{\theta}\xi_{\mu}\xi_{\nu}+p_{\varphi}\zeta_{\mu}\zeta_{\nu},
\label{finalfmag}
\eeqs
 while:
 \beqs
 \sigma&=&-4\rho^0\frac{B_0^2r^2\sin^2\theta}{\Lambda^3}.
\label{sigm}
 \eeqs
Here we defined $u_{\mu}=\left(-\sqrt{f(r)}\Lambda, 0, 0, 0\right)$ as the $4$-velocity of the fluid, while $\chi_{\mu}=(0, \Lambda/\sqrt{f(r)}, 0, 0)$, $\xi_{\mu}=(0, 0, r\Lambda, 0)$ and $\zeta_{\mu}=(0, 0, 0, \frac{r\sin\theta}{\Lambda})$ are respectively spacelike unit vectors in the radial and the transverse angular directions in the new geometry (\ref{ds2}).

Finally, $F=dA$ is the Maxwell field strength with the magnetic potential $A=A_{\varphi}d\varphi$, $J_{\mu}=(0, 0, 0, j_{\varphi})$ is the $4$-current that sources the electromagnetic field, while the stress-energy tensor of the electromagnetic field is defined as usual by (\ref{Temfield}).

The only non-vanishing component of the $4$-current $J_{\mu}$ which sources the electromagnetic field is:
\beqs
 J_{\varphi}&=&4\rho^0\frac{B_0r^2\sin^2\theta}{\Lambda^4}.
\label{Jmag}
 \eeqs
We explicitly checked that this is an exact solution of the Einstein-Maxwell-fluid equations (\ref{eomEMF}) using Maple.

In the followings, we shall focus on the special case of a magnetized Kiselev black hole surrounded by quintessential matter with $w=-\frac{2}{3}$, that is:
\begin{equation}
f(r) = 1 - \frac{2M}{r} - kr
\label{kisf}
\end{equation}
 where $k$ is denoted here as the quintessence parameter. 
In general, the Equation of State parameter $w$ must be in the range $-1 < w < -1/3$ in order to induce the Universe acceleration. However, theoretical studies mostly concentrates on the particular case with $w = -2/3$. This value is in the allowed range and also leads to a relative simplicity of treatment due to the linear contribution in the metric function. The solutions of the equation $f(r)=0$ are:
\begin{equation}
r_- = \frac{1 - \sqrt{1-8kM}}{2k} , \quad r_+ = \frac{1 + \sqrt{1-8kM}}{2k}
\label{rmp}
\end{equation}
where $k \leq 1/(8M)$. In the relation (\ref{rmp}), $r_ -$ is the black hole horizon while $r_+$ is an effective cosmological horizon located at $r_+ > r_-$.

For $B_0=0$, the relation (\ref{ds2}) with (\ref{kisf}) is the usual Kiselev line element, while for $k=0$, we recover the Ernst solution \cite{Ernst}, also known as the Schwarzschild–Melvin solution.

\section{Charged particles in the magnetized Kiselev solution}
\subsection{Equations of motion}

\begin{figure}[H]
\centering
\includegraphics[width=0.45\textwidth]{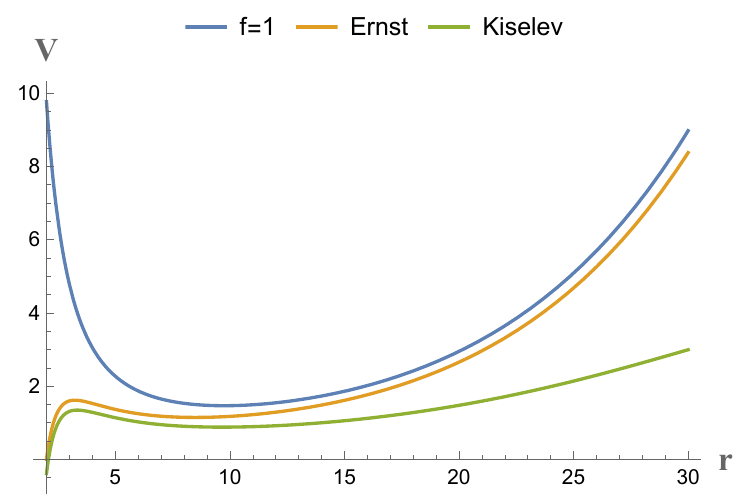} 
\hspace{0.2cm}
\includegraphics[width=0.45\textwidth]{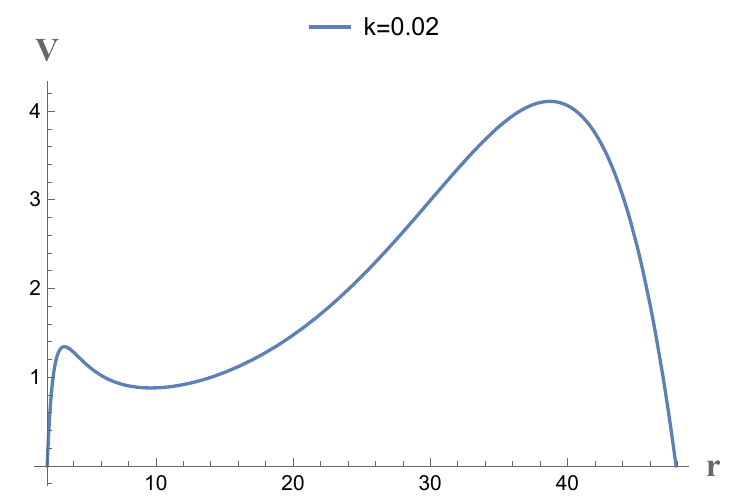} 
\caption{{\it Left panel}. 
The effective potential (\ref{Vmag}) in the equatorial plane for different metric functions. {\it Right panel}. The effective potential (\ref{Vmag}) in the equatorial plane for the metric function  (\ref{kisf}), with $k=0.02$ and $r \in [r_- , r_+ ]$. The numerical values of the other parameters are: $M=1$, $L=6$, $\varepsilon =1$, $B_0=0.04$.}
\label{PotEq}
\end{figure}

For a test particle with mass $m_0$ and electric charge $q$, it's motion can be described using the following Lagrangian:
\begin{equation}
{\cal L} = \frac{1}{2} \left \lbrace   - f(r) \Lambda^2  \dot{t}^2 + \frac{\Lambda^2}{f(r)}  \dot{r}^2
+ \Lambda^2 r^2 \dot{\theta}^2 + \frac{r^2 \sin^2 \theta}{\Lambda^2} \dot{\varphi}^2  \right \rbrace  + \frac{\varepsilon B_0 r^2 \sin^2 \theta}{\Lambda} \dot{\varphi},
\label{Lag}
\end{equation}
where we defined $\varepsilon = q/m_0$. Since $t$ and $\varphi$ are cyclical coordinates, one obtains the following constants of motion:
\begin{equation}
E \, = \, f(r)\Lambda^2 \dot{t} 
\label{E}
\end{equation}
and
\begin{equation}
L = \frac{r^2 \sin^2 \theta}{\Lambda^2} \dot{\varphi} + \frac{\varepsilon B_0 r^2 \sin^2 \theta}{\Lambda} \; ,
\label{L}
\end{equation}
Using these expressions one can cast the Euler--Lagrange equations for the coordinates $r$ and $\theta$ in the following form::
\begin{eqnarray}
\ddot{r}&  = & \left[ - \frac{\Lambda,_{r} }{\Lambda} + \frac{f^{\prime}}{2f} \right] \dot{r}^2 +
f r^2 \left[ \frac{\Lambda,_{r}}{\Lambda} + \frac{1}{r} \right] \dot{\theta}^2
- \frac{2 \Lambda,_{\theta}}{\Lambda} \dot{r} \dot{\theta}  - \frac{E^2}{\Lambda^4} \left[  \frac{\Lambda,_{r} }{\Lambda} + \frac{f^{\prime}}{2f} \right] \nonumber \\*
& &  - \frac{f}{r^2 \sin^2 \theta } 
\left[ \frac{\Lambda,_r}{\Lambda} - \frac{1}{r} \right] \left[ L - \frac{\varepsilon B_0 r^2 \sin^2 \theta}{\Lambda} \right]^2 - \frac{\varepsilon B_0 f}{\Lambda} \left[ \frac{\Lambda,_r}{\Lambda} - \frac{2}{r} \right] \left[ L - \frac{\varepsilon B_0r^2 \sin^2 \theta}{\Lambda} \right]
\label{rEq}
\end{eqnarray}
and
\begin{eqnarray}
\ddot{\theta}&  = & \frac{\Lambda,_{\theta}}{f r^2 \Lambda} \left[ \dot{r}^2 - \frac{E^2}{\Lambda^4} \right]
- \frac{\Lambda,_{\theta}}{\Lambda} \dot{\theta}^2
- 2 \left[  \frac{\Lambda,_{r} }{\Lambda} + \frac{1}{r} \right] \dot{r} \dot{\theta} - \frac{1}{r^4 \sin^2 \theta } 
\left[ \frac{\Lambda,_{\theta}}{\Lambda} - \cot \theta \right]   \left[ L - \frac{\varepsilon B_0r^2 \sin^2 \theta}{\Lambda} \right]^2 \nonumber \\*
& &  - \frac{\varepsilon B_0}{r^2 \Lambda}  \left[ \frac{\Lambda,_{\theta}}{\Lambda} - 2 \cot \theta \right]   \left[ L - \frac{\varepsilon B_0r^2 \sin^2 \theta}{\Lambda} \right]
\label{tEq}
\end{eqnarray}
Note also that we obtain the relation:
\begin{equation}
\dot{\varphi} =  \frac{\Lambda^2}{r^2 \sin^2 \theta} \left[ L - \frac{\varepsilon B_0 r^2 \sin^2 \theta}{\Lambda} \right] 
\label{phi}
\end{equation}
On the other hand, the four-velocity normalization relation
\begin{equation}
- f (r) \Lambda^2 \dot{t}^2  + \frac{\Lambda^2}{f (r)} \dot{r}^2 + \Lambda^2 r^2 \dot{\theta}^2  + \frac{r^2 \sin^2 \theta}{\Lambda^2} \dot{\varphi}^2   = -1
\label{taum}
\end{equation}
with the constants of motion (\ref{E}) and (\ref{L}), leads to the equation
\begin{equation}
\Lambda^4 \left[ \dot{r}^2 + f(r) r^2 \dot{\theta}^2 \right]   = E^2 -V_{eff} 
\label{rdot}
\end{equation}
The effective potential
\begin{equation}
V_{eff} =  f(r) \Lambda^2 \left[ 1 + \frac{\Lambda^2}{r^2 \sin^2 \theta} \left( L - \frac{\varepsilon B_0 r^2 \sin^2 \theta}{\Lambda} \right)^2 \right] ,
\label{Vmag}
\end{equation}
in the equatorial plane, is represented in the left panel of the figure \ref{PotEq}, for $f=1$ (the Melvin case), for $f(r) = 1-2M/r$ (the Ernst spacetime) and for the metric function (\ref{kisf}). Depending on the particle's energy and on the starting point, one may have different types of orbits.
These can be obtained by numerically integrating the equations (\ref{rEq}), (\ref{tEq}) and (\ref{phi}), for specific orbital parameters.

\subsection{Particle trajectories in Melvin spacetime}

Let us start with the case of the Melvin magnetic universe which has been largely discussed in \cite{lim}. In addition to the Lorentz interaction, the magnetic field exerts a gravitational force on the test particle. According to the author, there is not a significant difference to bound orbits in the absence of the black hole aside from a weaker gravitational force.  Although some of the next results may be repetitive as done in previous work \cite{lim}, they are given for comparison.

With $f=1$, the equations of motion (\ref{rEq}) and (\ref{tEq}) turn into the simpler expressions
\begin{eqnarray}
\ddot{r}&  = & - \frac{\Lambda,_{r} }{\Lambda} \left[ \dot{r}^2  + \frac{E^2}{\Lambda^4}  \right] +
r^2 \left[ \frac{\Lambda,_{r}}{\Lambda} + \frac{1}{r} \right] \dot{\theta}^2
- \frac{2 \Lambda,_{\theta}}{\Lambda} \dot{r} \dot{\theta}  \nonumber \\*
& &  - \frac{1}{r^2 \sin^2 \theta } 
\left[ \frac{\Lambda,_r}{\Lambda} - \frac{1}{r} \right] \left[ L - \frac{\varepsilon B_0 r^2 \sin^2 \theta}{\Lambda} \right]^2 - \frac{\varepsilon B_0}{\Lambda} \left[ \frac{\Lambda,_r}{\Lambda} - \frac{2}{r} \right] \left[ L - \frac{\varepsilon B_0r^2 \sin^2 \theta}{\Lambda} \right]
\label{rEqM}
\end{eqnarray}
and
\begin{eqnarray}
\ddot{\theta}&  = & \frac{\Lambda,_{\theta}}{r^2 \Lambda} \left[ \dot{r}^2 - \frac{E^2}{\Lambda^4} - r^2  \dot{\theta}^2 \right]
- 2 \left[  \frac{\Lambda,_{r} }{\Lambda} + \frac{1}{r} \right] \dot{r} \dot{\theta} - \frac{1}{r^4 \sin^2 \theta } 
\left[ \frac{\Lambda,_{\theta}}{\Lambda} - \cot \theta \right]   \left[ L - \frac{\varepsilon B_0r^2 \sin^2 \theta}{\Lambda} \right]^2 \nonumber \\*
& &  - \frac{\varepsilon B_0}{r^2 \Lambda}  \left[ \frac{\Lambda,_{\theta}}{\Lambda} - 2 \cot \theta \right]   \left[ L - \frac{\varepsilon B_0r^2 \sin^2 \theta}{\Lambda} \right]
\label{tEqM}
\end{eqnarray}
and the effective potential (\ref{Vmag}) becomes:
\begin{equation}
V_{eff}(r,\theta)=\Lambda^2\left[1+\frac{\Lambda^2}{r^2\sin^2\theta}\left(L-\frac{\varepsilon B_0 r^2 \sin^2\theta}{\Lambda}\right)^2\right]
\end{equation}

\begin{figure}[H]
\centering
\includegraphics[scale=0.4,trim = 2cm 12cm 2cm 2cm]{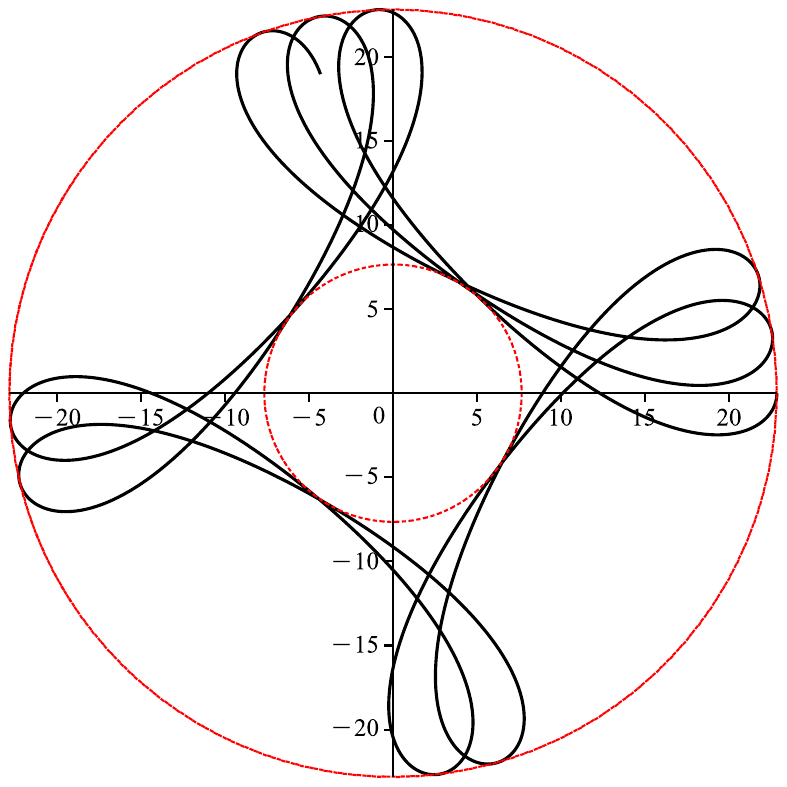}
\includegraphics[scale=0.4,trim = 2cm 12cm 2cm 2cm]{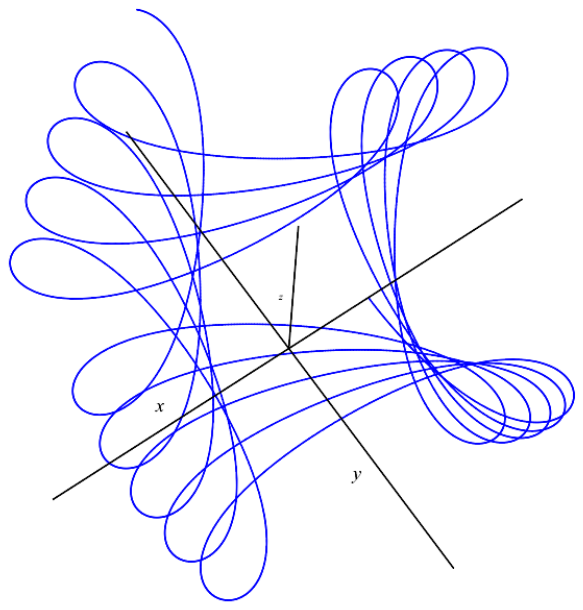}
\caption{{\it Left panel.} Polar plot of a bound orbit in the equatorial plane. {\it Right panel.}  3D plot of orbit. The numerical values are: $\varepsilon=1$, $B_0=0.02$, $L=6$ and $E^2=1.22$. }
\label{Mel1}
\end{figure}

\begin{figure}[H]
\centering
\includegraphics[scale=0.4,trim = 2cm 12cm 2cm 2cm]{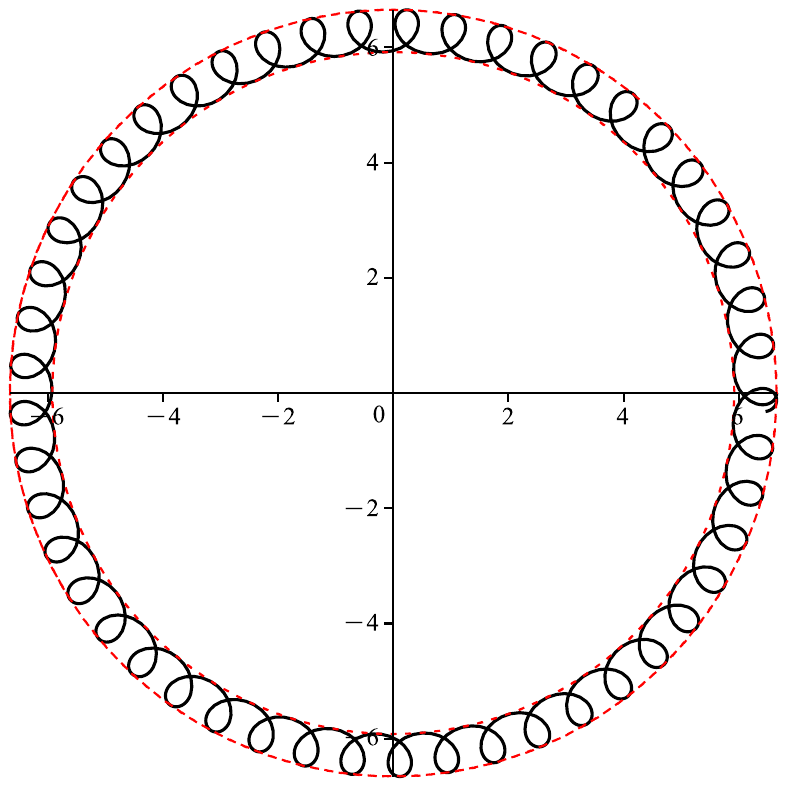}
\includegraphics[scale=0.4,trim = 2cm 12cm 2cm 2cm]{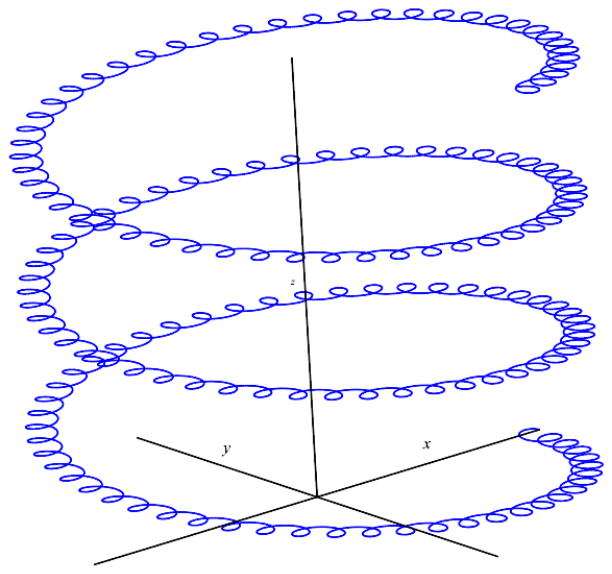}
\caption{{\it Left panel.} Polar plot of a bound orbit in the equatorial plane. {\it Right panel.}  3D plot of orbit. The numerical values are: $\varepsilon=1$, $B_0=0.025$, $L=5$ and $E^2=1.06$. }
\label{Mel2}
\end{figure}

A numerical integration of the equations (\ref{rEqM}) and (\ref{tEqM}) allows us to display the polar plot of the bounded orbits of the test particle for different orbital parameters (see the figures \ref{Mel1} and \ref{Mel2}). 


As it can be noticed in the figure \ref{Mel2}, the particle can be confined to a bound orbit in the equatorial plane but under a small perturbation, the particle's trajectory leaves the equatorial plane and describes a helicoid along the $z$ axis. 

When the metric function in (\ref{ds2}) describes a Schwarzschild BH, one deals with the Ernst spacetime \cite{Ernst} and the whole analysis becomes more complex. As it is known, the Ernst solution is not asymptotically flat and, for large radial distances, it possess the characteristics of the Melvin spacetime. For an analytical investigation, one can take the range of the radial coordinate as being $r \gg2M$ and employ a perturbative approach \cite{Frolov}.
However, in the presence of quintessence, the linear term in the metric function (\ref{kisf}) is linearly increasing with $r$ and this term might play a significant contribution for values of $r$ close to the cosmological horizon. 

\subsection{Particle trajectories in the magnetized Kiselev BH}

\subsubsection{The effective potential and critical points}

\begin{figure}[H]
\centering
\includegraphics[scale=0.4,trim = 1cm 6cm 1cm 1cm]{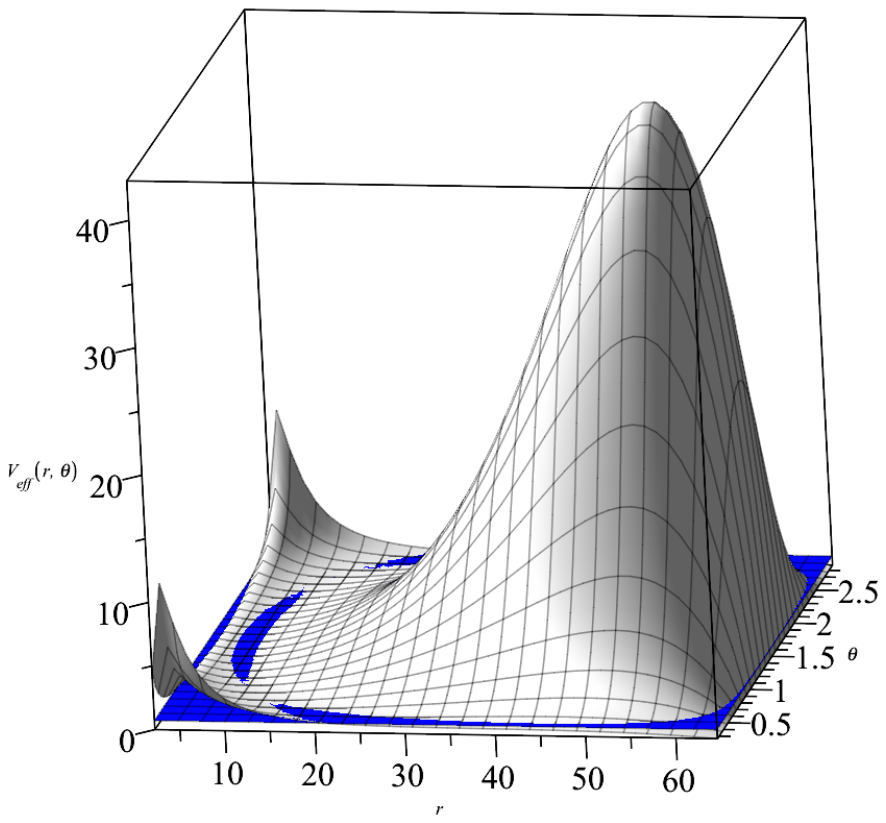}
\includegraphics[scale=0.4,trim = 1cm 10cm 1cm 1cm]{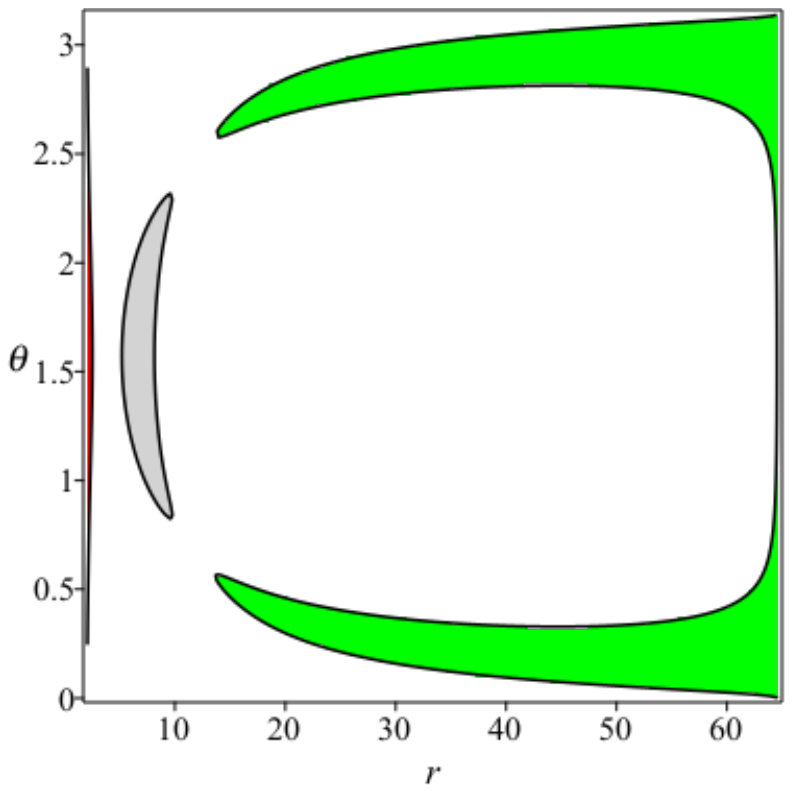}
\caption{{\it Left panel.} 3D plot of the effective potential (\ref{Vmag}), for $r \in [r_-, r_+]$ and $\theta \in [\pi/12, 11\pi/12]$. The blue plane represents the energy $E^2=0.715$. {\it Right panel.} Projection of the effective potential in the $(r,\theta)-$ plane. The shaded areas represent the allowed regions of motion where $E^2 \geq V(r,\theta)$. The black curves correspond to the solutions of the equation $E^2=V(r,\theta)$.  The numerical values are: $M=1$, $k=0.015$, $L=5$, $\varepsilon=3$, $B_0=0.03$. }
\label{Vmag3D}
\end{figure}

For the metric function (\ref{kisf}) which is the aim of the present work, the 3-dimensional particle's trajectory can be obtained by numerically integrating the general equations (\ref{rEq}), (\ref{tEq}) and (\ref{phi}). 
 Compared to the Melvin and Ernst cases, the effective potential (\ref{Vmag}) with the Kiselev metric function (\ref{kisf}) has a different behavior. It is defined for $r \in [r_- , r_+ ]$ and vanishes in $r_+$ and $r_-$, where $r_{\pm}$ are the horizons given in (\ref{rmp}). A graphical representation of the potential is given in the figure \ref{Vmag3D}, where the particle's energy $E^2$ is represented by the blue horizontal plane. The condition $E^2 =V_{eff}$ determines the allowed regions in the $( r, \theta)$-plane.  In the right panel of the figure \ref{Vmag3D} these are colored in red, green and gray.

The shape of the potential is strongly depending on the parameters values. 
The range of $B_0$ has been stated as being $B_0 \in [0.0085 , 0.085]$ \cite{Frolov, Esteban}. For $B_0 < 0.0085$, the situation is similar to the Schwarzschild case. As for the maximum value $B_0 = B_{max} \approx 0.085$, using the relation which gives the conversion between the dimensionless quantity $M B_0$ and $B_G$ expressed in Gauss \cite{Esteban} 
\[
M B_0=8.5 \times 10^{-21} \frac{M}{M_S} B_G,
\]
one may notice that $B_{max}$ can be reached only for supermassive magnetized black holes with $M = 10^{11} M_S$ and $B_G = 10^8$ (G). However, close to this maximum magnetic induction, no bounded orbits are allowed \cite{Esteban}. Together with the magnetic induction, the quintessence parameter $k \in [0, 1/(8M)]$ plays an important role. For $k=0$, we recover the Ernst spacetime, while for the maximum value $k=1/(8M)$, one has only one horizon in $4M$. As it can be noticed in figure \ref{Vmag3D}, for small values of $k$, there is a second maximum of the potential, close to the cosmological horizon which is most prominent for particles moving in the equatorial plane.
As $k$ increases, the second maximum vanishes and there is only one maximum, close to the black hole's horizon, which corresponds to an unstable circular orbit.

Once the model's parameters $M$, $B_0$ and $k$ are fixed, one may have different types of trajectories, depending on the particle's energy, specific charge and angular momentum. 
The critical points of the potential (\ref{Vmag}) can be found by imposing the conditions:
\begin{equation}
\frac{\partial V_{eff}}{\partial r}= 0 = \frac{\partial V_{eff}}{\partial \theta}
\end{equation}
which lead to the following equations in terms of the angular momentum $L$:
\begin{eqnarray}
& &
\Lambda^2\left[4fr \partial_r \Lambda +(f'r-2f)\Lambda\right]L^2-2\varepsilon B_0 \Lambda r^3\sin^2\theta(f'\Lambda+3f\partial_r \Lambda)L
\nonumber \\*
& &
+r^3\sin^2\theta\left[(\varepsilon^2B_0^2r^2\sin^2\theta+1)(f'\Lambda+2f\partial_r \Lambda)+2\varepsilon^2B_0^2fr\Lambda\sin^2\theta\right]=0
\label{Lr}
\end{eqnarray}
and
\begin{eqnarray}
& &
2f\Lambda^3\left(2\partial_{\theta}\Lambda \sin\theta-\Lambda\cos\theta\right)L^2-\left(6\varepsilon B_0 fr^2\Lambda^2\partial_{\theta}\Lambda\sin^3\theta\right) L
\nonumber \\*
& & +2fr^2\Lambda\sin^3\theta\left[(\varepsilon^2B_0^2r^2+1)\partial_{\theta}\Lambda+\varepsilon^2 B_0^2 \Lambda r^2\sin\theta\cos\theta\right]=0
\label{Ltheta}
\end{eqnarray}
On the equatorial plane, i.e. $\theta=\pi/2$, the left hand side of the equation (\ref{Ltheta}) is zero. However, in the case of  off-equatorial plane, one has to impose that the equations (\ref{Lr}) and (\ref{Ltheta}) have at least one common solution. Using Cramer's rule one finds that the radial coordinate of an off-equatorial critical point is located at
\begin{equation}
r_* =\sqrt{\frac{2M}{k}}
\label{roff}
\end{equation}
and it is independent of the parameters $\varepsilon$, $L$ and $B_0$. The existence of such a point is solely due to the presence of quintessence. For instance, in Ernst spacetime, all critical points are located in the equatorial plane. 
Using the relations (\ref{LambdaB}) and  (\ref{roff}), the equation (\ref{Ltheta}) becomes
\begin{equation}
24 B_0^4 M^3(\varepsilon-B_0 L)^2\sin^6\theta+4 B_0^2 M^2 k(5 B_0^2 L^2-6\varepsilon B_0L+\varepsilon^2+2)\sin^4\theta+2B_0 L^2 Mk^2\sin^2\theta-k^3L^2=0
\label{eqtheta}
\end{equation}
where $\theta$ represents the polar coordinate corresponding to the off-equatorial plane critical point.  Due to the axial symmetry, there are two such points symmetric with respect to the $\pi/2$ plane. For specific values of the parameters, one has to solve the cubic equation (\ref{eqtheta}) in $\sin^2 \theta$ and identify the real positive solutions in the physical range $[0,1]$. These correspond to the off-equatorial plane critical points. 
One may check the nature of the critical point located at $r_* =\sqrt{2M/r}$ and $\theta_* \neq \pi/2$ by evaluating the corresponding hessian.

As an example, let us consider the effective potential represented in the figure \ref{Vmag3D}. This has three extreme points (one minimum and two maxima), all located in the equatorial plane and two saddle points located outside the equatorial plane. These points are classified in the table \ref{tab1}, with the corresponding radial and polar coordinates and the values of the potential. 
\\
\begin{table}[h]
\centering
\begin{tabular}{|c|c|c|c|}
  \hline
 \it{Nature of the critical point} & \it{Radial coordinate $r$} & \it{Polar coordinate $\theta$} & \it{The potential value} \\ \hline
  saddle  & $11.55$ & $0.68$ & $V_s=0.720$ \\ \hline
  saddle & $11.55$  & $2.25$ & $V_s=0.720$ \\ \hline
  local maximum & $3.21$  & $\pi/2$ & $V_{\max1}=0.887$ \\ \hline
  local minimum  & $6.60$ & $\pi/2$ & $V_{\min}=0.670$ \\ \hline
  local maximum  & $53.67$ & $\pi/2$ & $V_{\max2}=43.164$ \\ \hline
\end{tabular}
\caption{Classification of the critical points of the potential represented in the figure \ref{Vmag3D}.}
\label{tab1}
\end{table}

The intersection of the potential surface with the blue horizontal plane corresponding to the particle's energy is leading to the allowed regions represented in the right panel. Depending on the energy and initial conditions, the particle may have different types of trajectories as it can be seen below.

\subsubsection{Classification of orbits}

\begin{itemize}
  \item[$\bullet$] {\it Capture orbits.} The particle with $E^2 \in ( 0 , V_{max1} )$, starting its journey in the red region in the right panel of the figure \ref{Vmag3D} will be captured by the black hole. The same happens for large values of the energy, $E^2 > V_{max1}$, for which the regions red, gray and green merge. Such a trajectory is given in the figure \ref{capture3D}.
  \begin{figure}[H]
\centering
\includegraphics[scale=0.4,trim = 1cm 6cm 1cm 1cm]{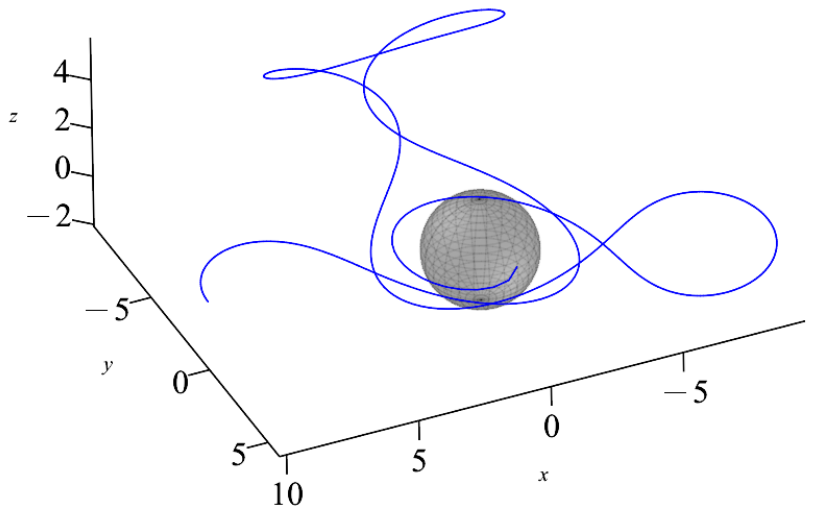}
\includegraphics[scale=0.4,trim = 1cm 10cm 1cm 1cm]{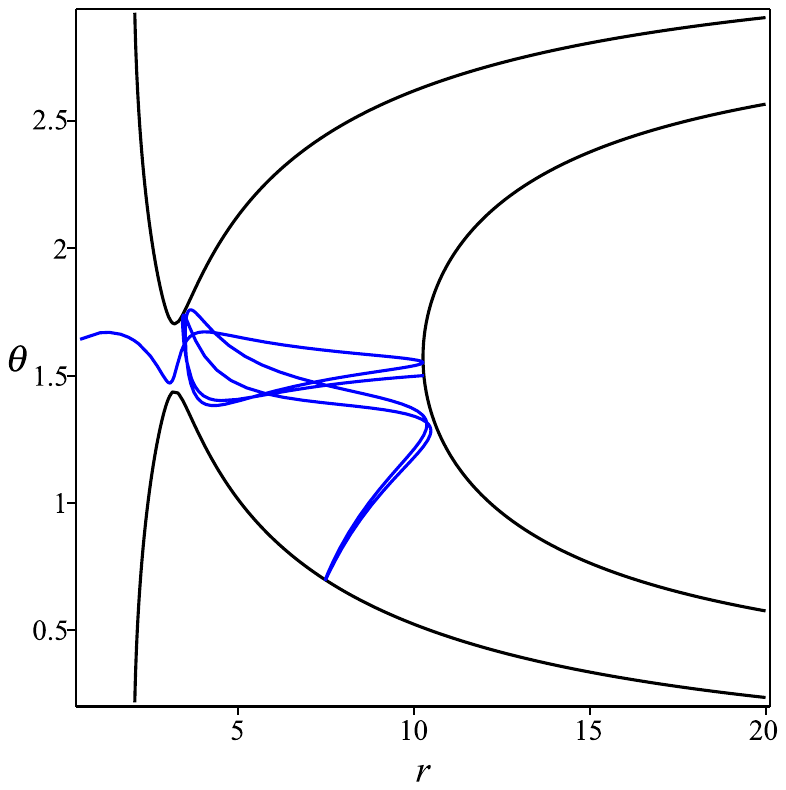}
\caption{{\it Left panel.} Capture orbit of a charged particle with $E^2 =0.900$. {\it Right panel.} Representation in $(r,\theta)-$plane.}
\label{capture3D}
\end{figure}
\item[$\bullet$] {\it Periodic bound orbits.} These are for particles with $V_{min} \leq E^2 < V_s$. For example, the particle with $E^2 = 0.715$ which starts its journey in the gray region (see the right panel of figure \ref{Vmag3D}) has the bounded curly trajectory represented in the figure \ref{bound3D1}. Once the particle's energy is decreasing towards the local minimum value $V_{min}$, the gray region is shrinking and the corresponding trajectory is approaching a circular orbit, as it can be noticed in the figure \ref{bound3D2}.
\begin{figure}[H]
\centering
\includegraphics[scale=0.4,trim = 1cm 6cm 1cm 1cm]{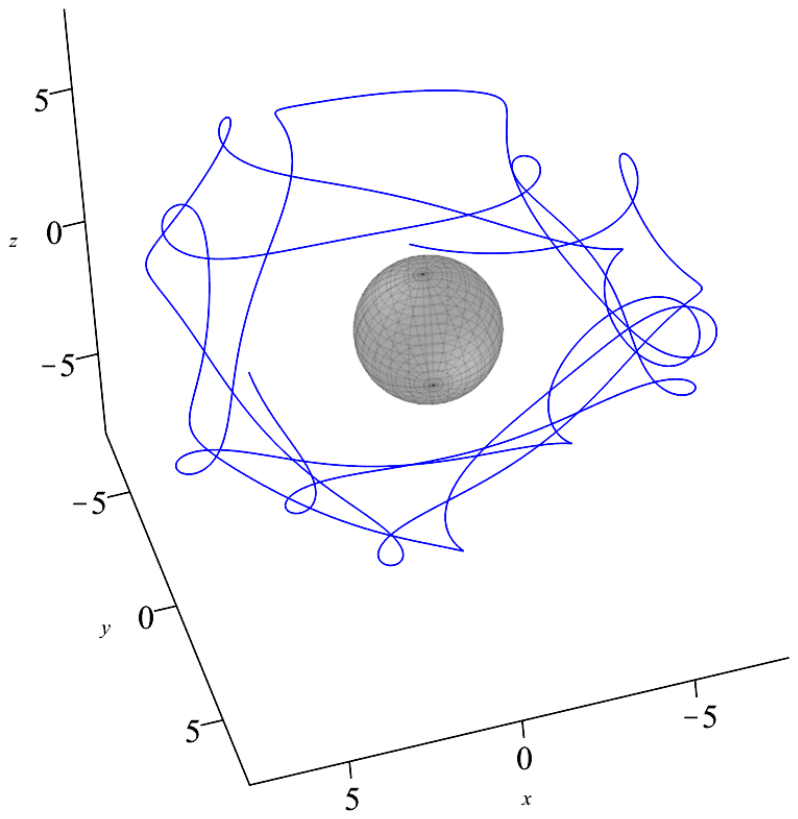}
\includegraphics[scale=0.4,trim = 1cm 10cm 1cm 1cm]{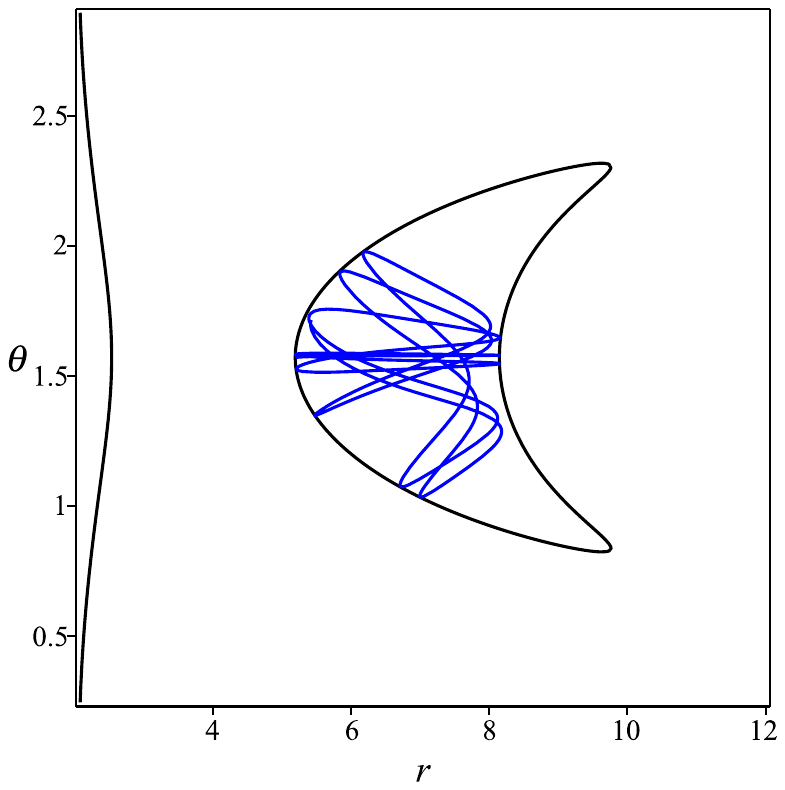}
\caption{{\it Left panel.} Bound orbit of a charged particle with $E^2 =0.715$ moving in the gray region. {\it Right panel.} Representation of the orbit in the $(r,\theta)-$ plane. }
\label{bound3D1}
\end{figure}
\begin{figure}[H]
\centering
\includegraphics[scale=0.4,trim = 1cm 6cm 1cm 1cm]{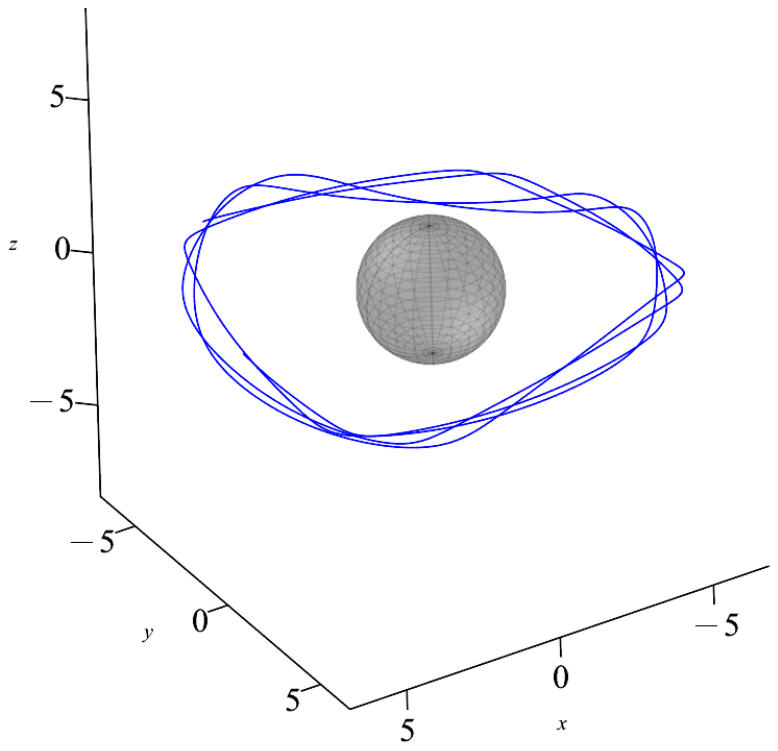}
\includegraphics[scale=0.4,trim = 1cm 10cm 1cm 1cm]{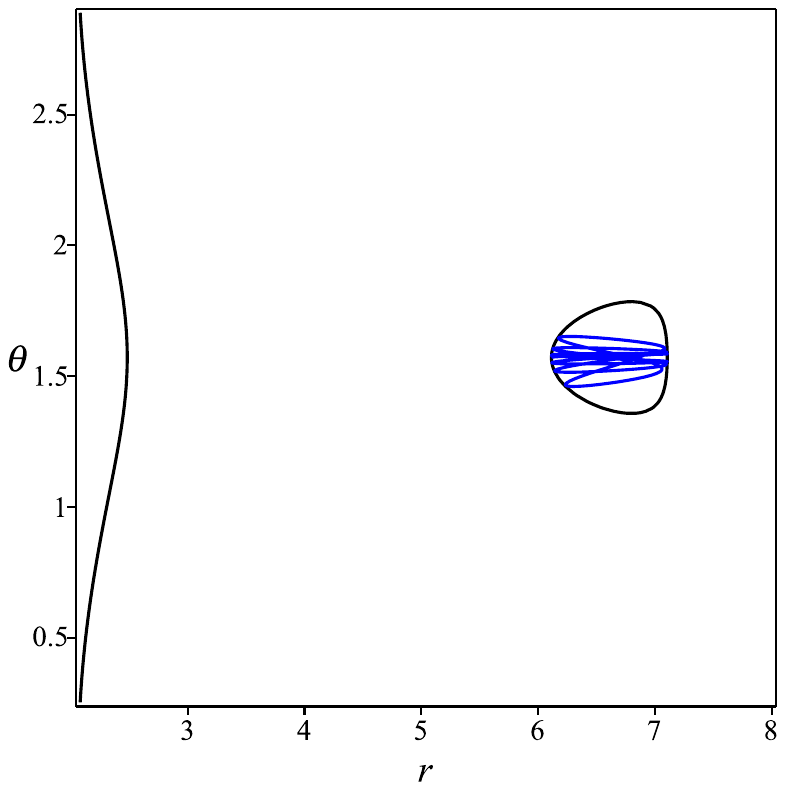}
\caption{{\it Left panel.} Bound orbit of a charged particle with $E^2 =0.675$. {\it Right panel.} Representation of the orbit in the $(r,\theta)-$ plane. }
\label{bound3D2}
\end{figure}
\item[$\bullet$] {\it Escape orbits.} The particle with $E^2 = 0.715$ starting its motion in the green region (see the right panel of figure \ref{Vmag3D}) will escape toward the cosmological horizon, as it can be seen in the figure \ref{escape1}. Once the energy increases and $E^2 \in ( V_s , V_{max1})$, the green region merges with the gray one and the particle is always escaping (see the figure \ref{escape2}). For $E^2 > V_{max1}$, all the allowed regions merge and the particle can either be captured or it escapes.
\begin{figure}[H]
\centering
\includegraphics[scale=0.4,trim = 1cm 6cm 1cm 1cm]{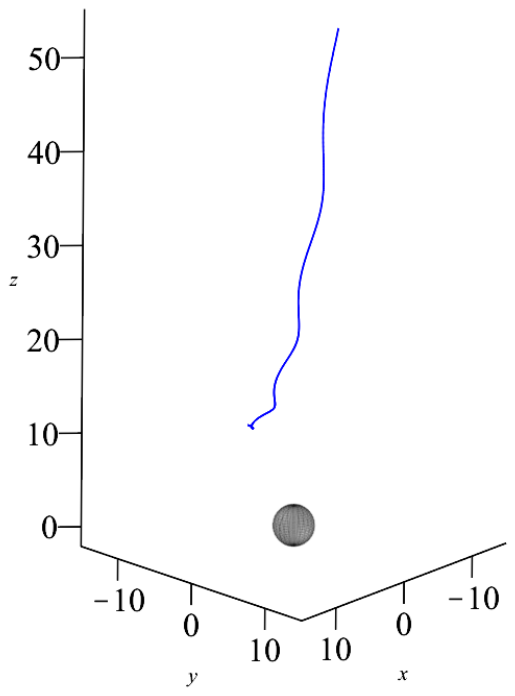}
\includegraphics[scale=0.4,trim = 1cm 10cm 1cm 1cm]{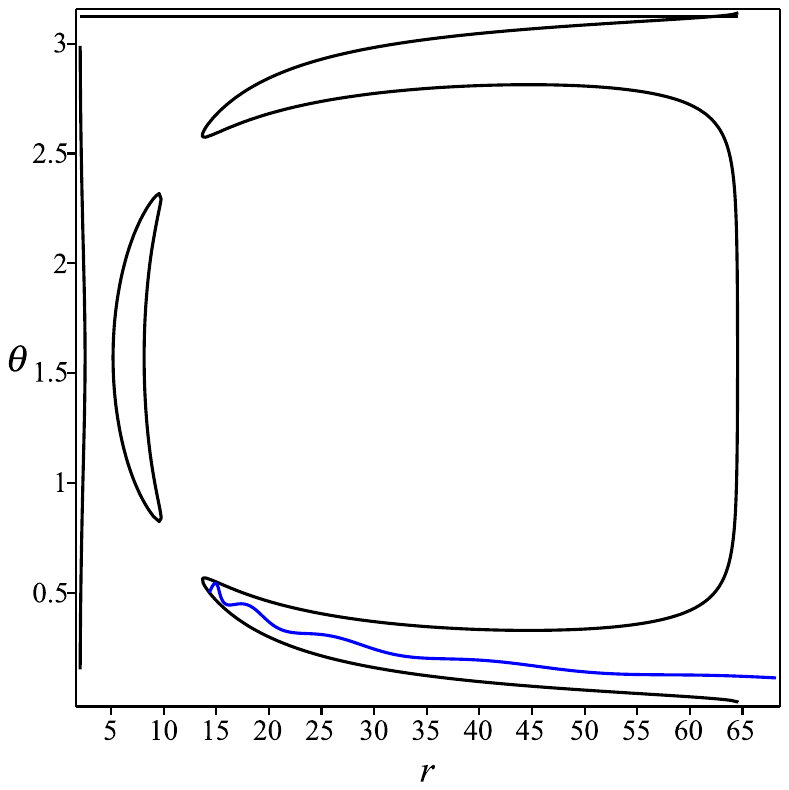}
\caption{{\it Left panel.} Escape orbit of a charged particle with $E^2 =0.715$ moving in the green region. {\it Right panel.} Representation in $(r,\theta)-$plane.}
\label{escape1}
\end{figure}
\begin{figure}[H]
\centering
\includegraphics[scale=0.4,trim = 1cm 6cm 1cm 1cm]{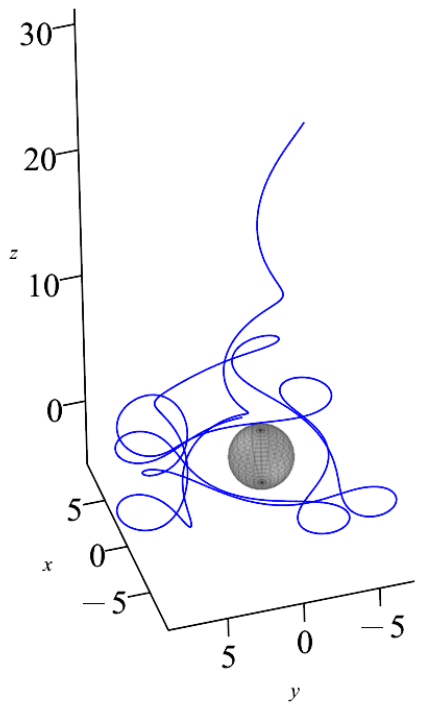}
\includegraphics[scale=0.4,trim = 1cm 10cm 1cm 1cm]{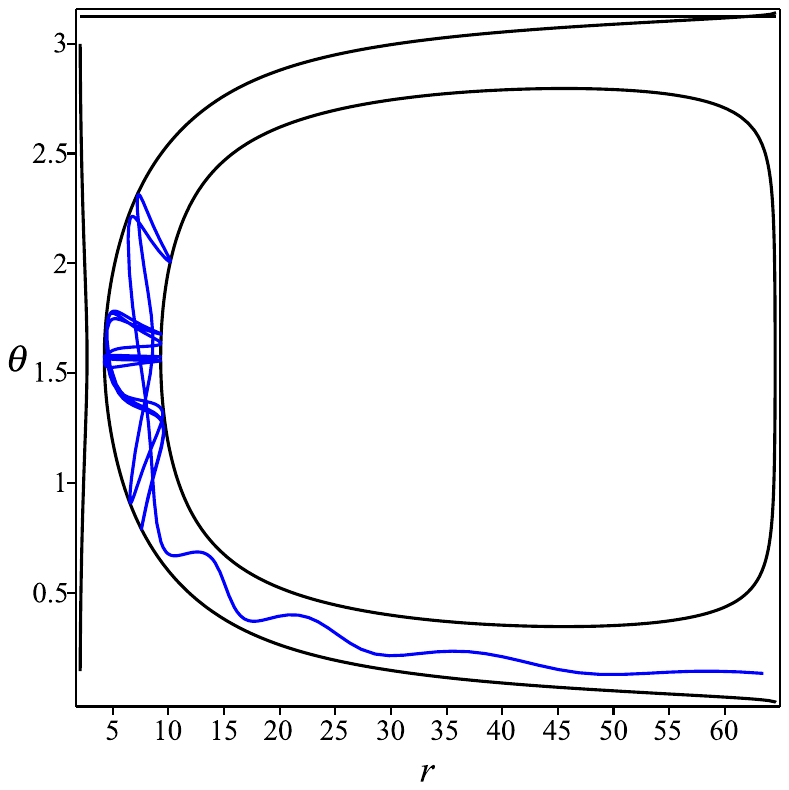}
\caption{{\it Left panel.} Escape orbit of a charged particle with $E^2 =0.800$. {\it Right panel.} Representation in $(r,\theta)-$plane.}
\label{escape2}
\end{figure}

\end{itemize}

An important characteristic which will be discussed in the following sections is that the particle's trajectory can
curl up into a cycloidlike motion \cite{lim}. 

\begin{figure}[H]
\centering
\includegraphics[width=0.45\textwidth]{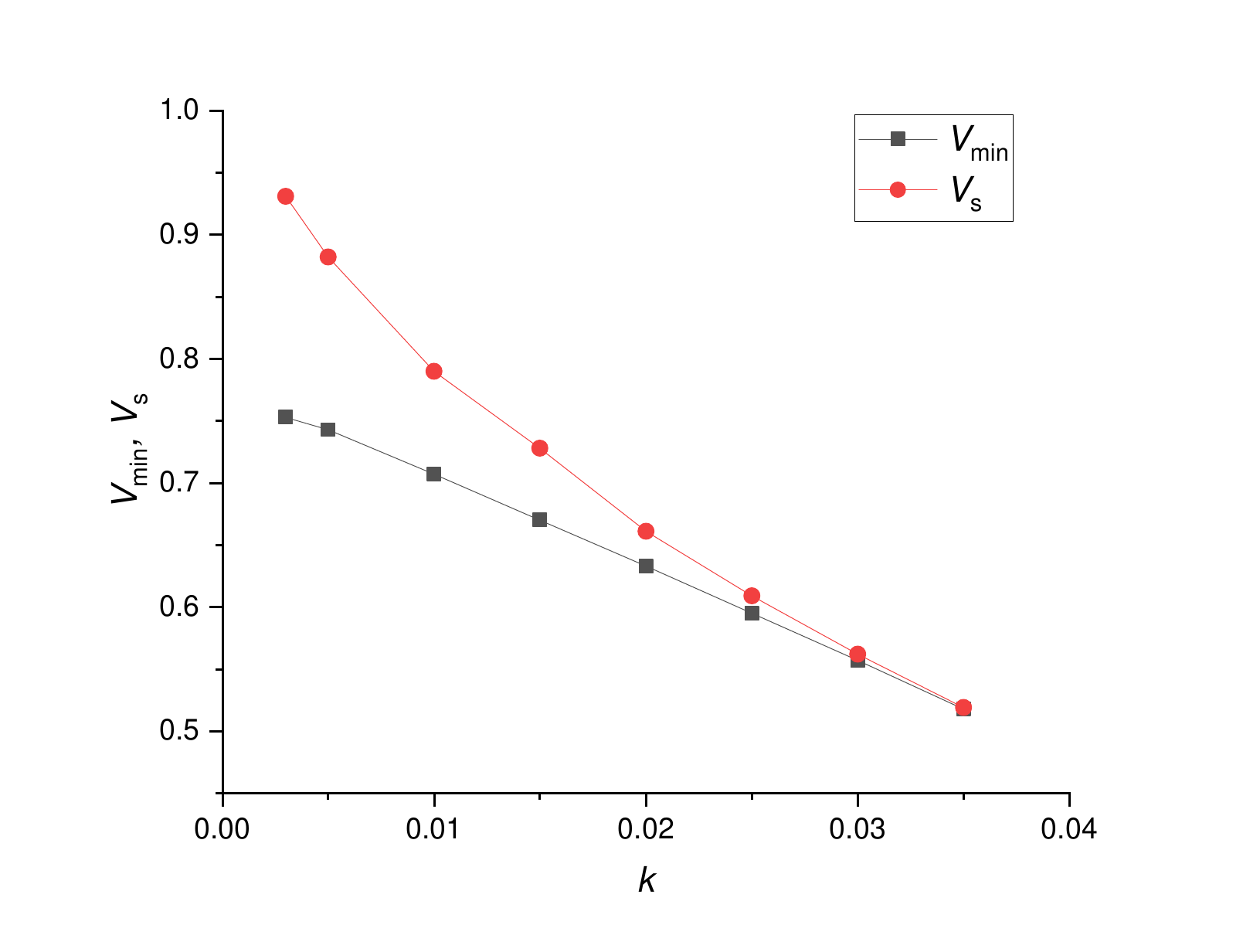}
\includegraphics[width=0.5\textwidth]{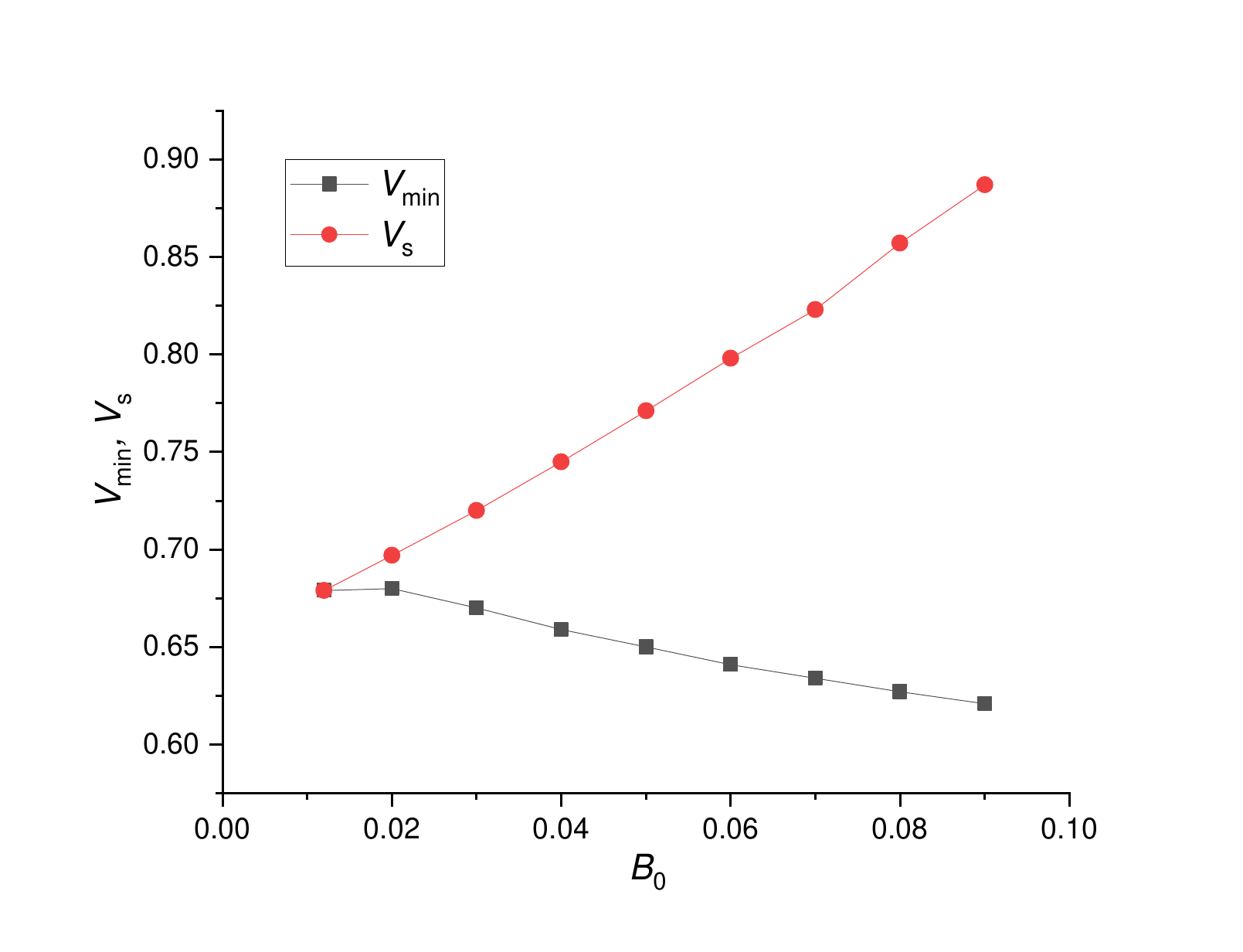}
\caption{{\it Left panel.} The plots of $V_{min}$ and $V_s$ for $k \in [0.003 , 0.035]$. The other numerical values are: $M=1$, $L=5$, $\varepsilon=3$, $B_0=0.03$. 
{\it Right panel.} The plots of $V_{min}$ and $V_s$ for different $B_0$ values. The other numerical values are: $M=1$, $L=5$, $\varepsilon=3$, $k=0.$. }
\label{VkB}
\end{figure}

Finally, let us discuss the influence of the model parameters $k$ and $B_0$ on the existence of bounded orbits. These occur as long as the particle's energy is in the range $E^2 \in ( V_{min} , V_s)$, where $V_{min}$ and $V_s$ are computed for the potential (\ref{Vmag}), once we fix the numerical values of the parameters. As we shall see, the parameters $k$ and $B_0$ compete against each other. In the left panel of the figure \ref{VkB}, we are giving the graphics of $V_{min}$ and $V_s$ as a function of $k$, for the numerical values used in the figure \ref{Vmag3D}.
One may notice that, as $k$ is increasing from $k=0.003$ to $k=0.035$, the range of $E^2$ shrinks. For $k >0.035$, there are no bounded trajectories and the particle is either captured or it is escaping. 

In the case of a fixed $k$ and increasing values of $B_0$, the range $E^2$ is increasing. For each value of $k$, there is maximum value of $B_0$ above which there are no bounded orbits.

\subsection{Types of orbits in the equatorial plane}

\subsubsection{The effective potential in the equatorial plane}
 
Since the magnetic field is assumed to be aligned axially, the geometry of the spacetime is axisymmetric and
the component of the angular
momentum in the $z$ direction is conserved. Similarly to most of the papers devoted to Melvin or Ernst spacetimes, in the followings, we shall consider $\theta = \pi/2$.
Indeed, for $\theta = \pi/2$ and $\dot{\theta}=0$, we have $\ddot{\theta} =0$ in (\ref{tEq}) and the motion is
confined to the equatorial plane. Thus, the $r-$equation can be solved by direct integration.
Outside the equatorial plane, the equations of motion can not be separated and solved since there is no general analytical
method to find all constants of motion.

\begin{figure}[H]
\centering
\includegraphics[width=0.4\textwidth]{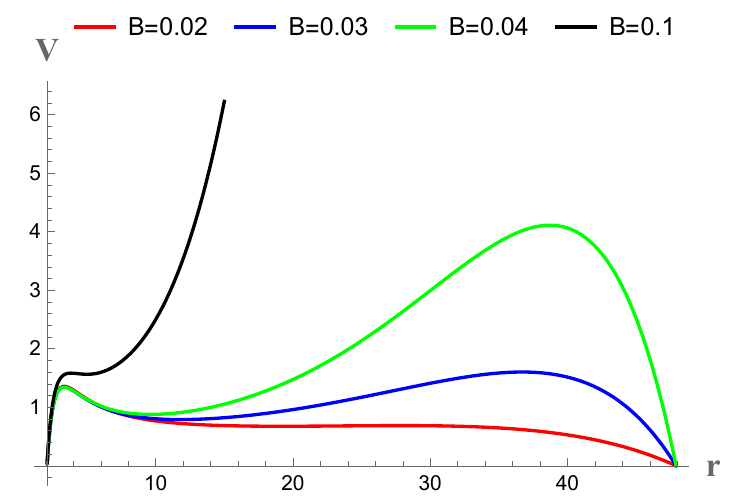}
\includegraphics[width=0.4\textwidth]{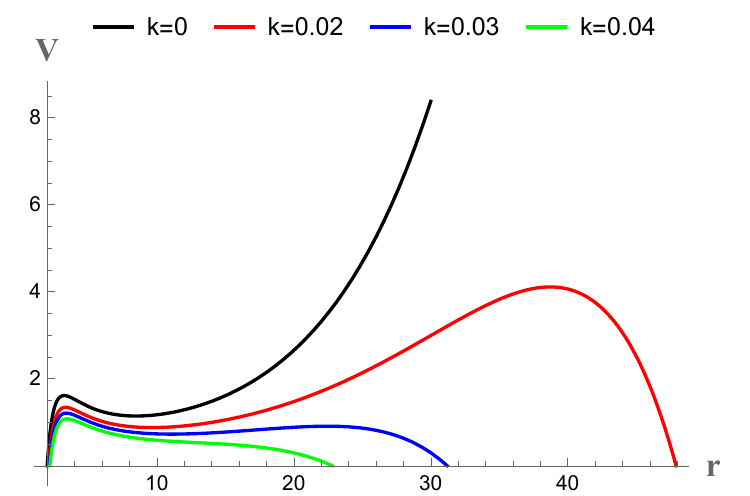}
\caption{{\it Left panel.} The effective potential (\ref{pot0}) for $k=0.02$ and different values of $B_0$. {\it Right panel.} The effective potential (\ref{pot0}) for $B_0 =0.04$ and different values of $k$. The other numerical values are: $M=1$, $\varepsilon=1$ and $L=6$.}
\label{PotEq1}
\end{figure}

\begin{figure}[H]
\centering
\includegraphics[scale=0.4,trim = 2cm 12cm 0cm 1cm]{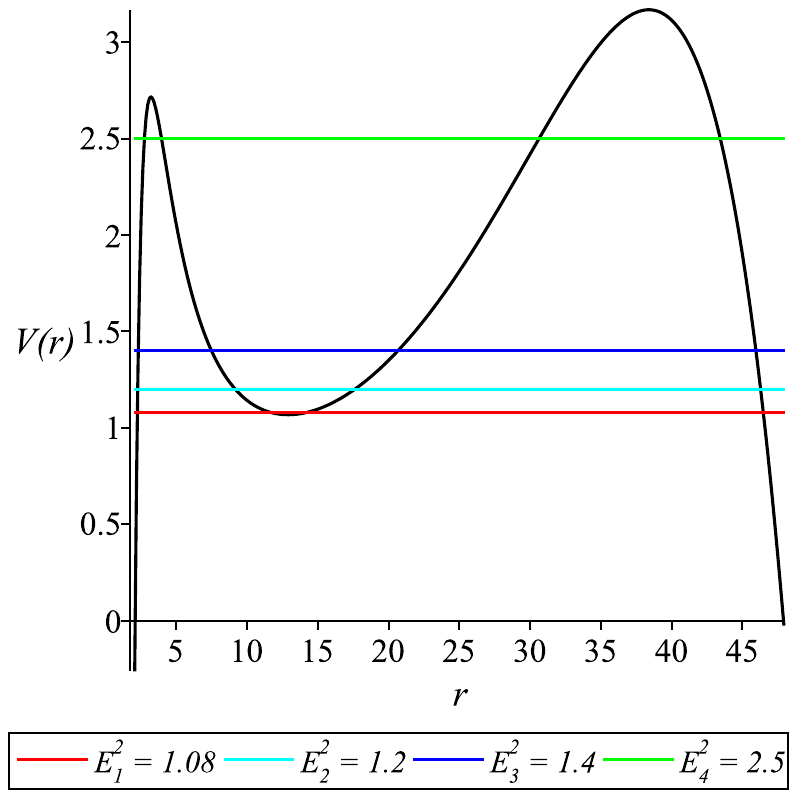}
\includegraphics[scale=0.4,trim = 2cm 12cm 0cm 1cm]{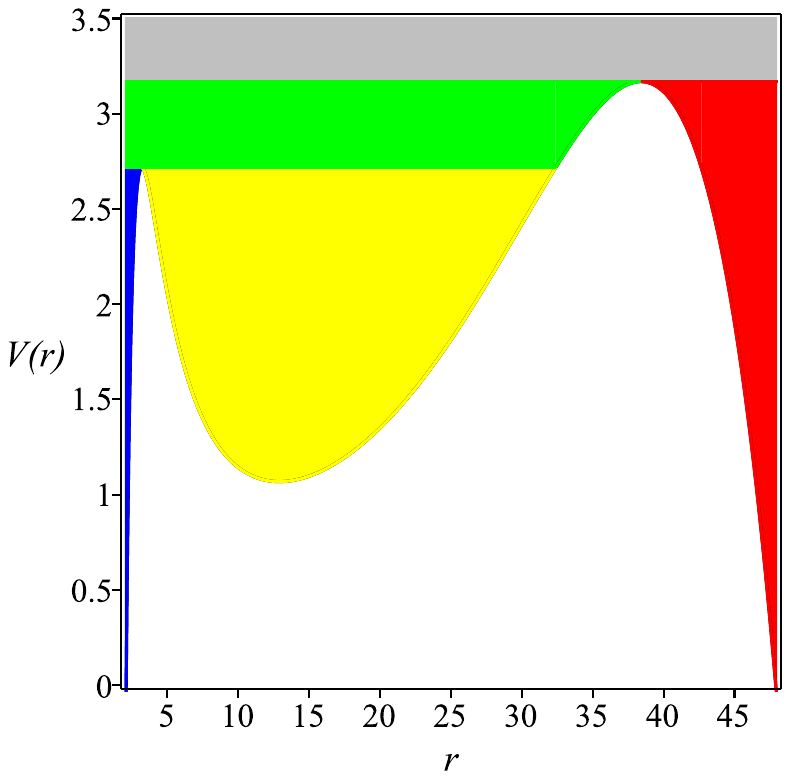}
\caption{{\it Left panel.} The effective potential (\ref{pot0}), with: $M=1$, $k=0.02$, $\varepsilon=1$, $B=0.04$ and $L=9$. The horizontal lines correspond to different values of $E^2$. {\it Right panel.} The effective potential (\ref{pot0}) and the physical regions corresponding to $E^2 \geq V(r)$. }
\label{PotEq2}
\end{figure}

For $\theta=\pi/2$, the function $\Lambda$ becomes:
\begin{equation}
\Lambda_0=1+B_0^2 r^2
\label{Lambda0}
\end{equation}
while the equations (\ref{phi}) and (\ref{rEq}) have the expressions
\begin{equation}
\dot{\varphi}=\frac{\Lambda_0^2}{r^2}\left(L-\frac{\varepsilon B_0 r^2 }{\Lambda_0}\right)
\label{varphi0}
\end{equation}
and
\begin{eqnarray}
& &
\ddot{r}=\left(\frac{f'}{2f}-\frac{ \Lambda_0'}{\Lambda_0}\right)\dot{r}^2-\frac{E^2}{\Lambda_0^4}\left(\frac{f'}{2f}+\frac{ \Lambda_0'}{\Lambda_0}\right)+\frac{f}{r^3}\left(1-\frac{r\Lambda_0'}{\Lambda_0}\right)\left(L-\frac{\varepsilon B_0 r^2 }{\Lambda_0}\right)^2
\nonumber \\*
& &
+ \frac{f\varepsilon B_0}{\Lambda_0 r}\left(2-\frac{r  \Lambda_0'}{\Lambda_0}\right)\left(L-\frac{\varepsilon B_0 r^2 }{\Lambda_0}\right)
\end{eqnarray}
where $f$ is the metric function (\ref{kisf}). The equation (\ref{rdot}) becomes
\begin{equation}
\Lambda^4_0 \dot{r}^2  = E^2 -V (r)
\end{equation}
where
\begin{equation}
V (r)=f\Lambda_0^2\left[1+\frac{\Lambda_0^2}{r^2}\left(L-\frac{\varepsilon B_0 r^2 }{\Lambda_0}\right)^2\right]
\label{pot0}
\end{equation}
The shape of the potential (\ref{pot0}) is strongly depending on the model's parameters $B_0$ and $k$ which compete against each other.
There are ranges of these parameters for which the effective potential has a minimum value between two maxima.
Thus, the particle with suitable energy can be trapped in bound orbits. As it can be noticed in the left panel of figure \ref{PotEq1}, as $B_0$ increases, the second maximum of the potential increases as well. For $B_0 > 0.1$, bounded trajectories are allowed only for very high values of angular momentum \cite{Esteban}. 
On the other hand, bound orbits exist only for low values of the quintessence parameter (see the right panel in the figure \ref{PotEq1}). The Ernst spacetime corresponds to $k=0$. This has only one event horizon in $r_h =2M$ and, due to the presence of the magnetic field, the metric is not asymptotically flat. Once the parameters $k$ comes into place, a second (cosmological) horizon appears on which the effective potential is vanishing. As $k$ increases,  the two horizons come closer to each other and the second maximum of the potential decreases until it disappears. At $k = k_{max} = 1/(8M)$, there is only one horizon, at $r=4M$.
In our further investigation, we will choose the values of $B_0$ and $k$ for which the effective potential allows bound orbits. An example is given in the figure \ref{PotEq2}.


The circular orbits can be obtained by imposing to the conditions: $V (r _0) = E^2$ and $V^{\prime}  (r_0) = 0$. If $V^{\prime \prime} (r_0) >0$ the circular orbit is stable, while for $V^{\prime \prime}  (r_0) < 0$, it is unstable. For a charged particle moving on a circular orbit of radius $r_0$, the relation $V^{\prime}  (r_0) = 0$ leads to the angular momentum's equation
\begin{eqnarray}
& &
\Lambda_0^2 \left[4f r_0 \Lambda_0^{\prime} +(f^{\prime} r_0-2f)\Lambda_0 \right]L^2-2 \varepsilon B_0 r_0^3 \Lambda_0 (3f \Lambda_0^{\prime} +f'\Lambda_0 )L \nonumber \\*
& & +r_0^3 \left[(2f\Lambda_0^{\prime}+f^{\prime} \Lambda_0 )(1+\varepsilon^2 B_0^2 r_0^2)+2 \varepsilon^2 B_0^2 fr_0 \Lambda_0 \right]=0
\nonumber
\end{eqnarray}
and the corresponding energy is $E^2 = V(r_0)$. For the potential represented in the figure \ref{PotEq2}, the particle may have two unstable circular orbits corresponding to the maxima $V_{max1} = 2.72$ and $V_{max2} = 3.17$ and a stable circular orbit corresponding to the minimum value of the effective potential $V_{min}=1.07$.  
When $k$ increases (see the right panel of figure \ref{PotEq1}), the second maximum vanishes and the potential has only one maximum, close to the black hole's horizon. Depending on the starting point, the particle moving on this unstable circular orbit can either fall into the black hole or reach the cosmological horizon.

\subsubsection{Classification of orbits}

The potential (\ref{pot0}) represented in the figure \ref{PotEq2} allows different types of orbits. In the right panel, we are using different colors for the physical regions where $E^2 \geq V(r)$. Thus, depending on the particle's energy and starting point $r_0$, the orbit can be: bound, escaping or capturing. 

\begin{itemize}
\item{} {\it Periodic bound orbits.}
In the yellow region, where $V_{min} \leq E^2 < V_{max1} < V_{max2}$, the particles are moving on stable bound orbits, neither falling into the black hole nor escaping to the cosmological horizon. The trajectory has two turning points, solutions of the equation $E^2=V(r)$, denoted by $r_1$ and $r_2$. Some examples of bounded orbits are plotted in the figures \ref{bound1} and \ref{bound2}, for different values of particle's energy represented in the left panel of the figure \ref{PotEq2} by horizontal lines. The turning points are represented by dashed red circles. The inner blue circle is for the black hole horizon. Thus, for $E^2 =1.08$ (the red horizontal line) close to $V_{min} =1.07$, the particle's trajectory is almost circular (see the left panel of the figure \ref{bound1}). As the energy is increasing, the two turning points move away from each other and this has a strong influence on the shape of the trajectory, as it can be noticed in the figures \ref{bound1} and \ref{bound2}. 

\begin{figure}[H]
\centering
\includegraphics[scale=0.4,trim = 2cm 12cm 0cm 1cm]{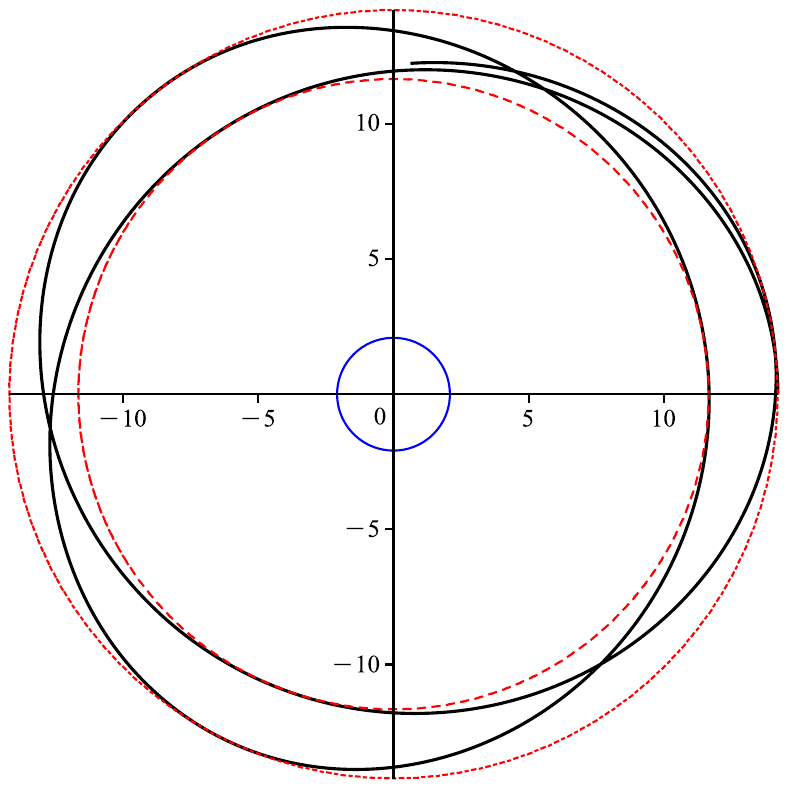}
\includegraphics[scale=0.4,trim = 2cm 12cm 0cm 1cm]{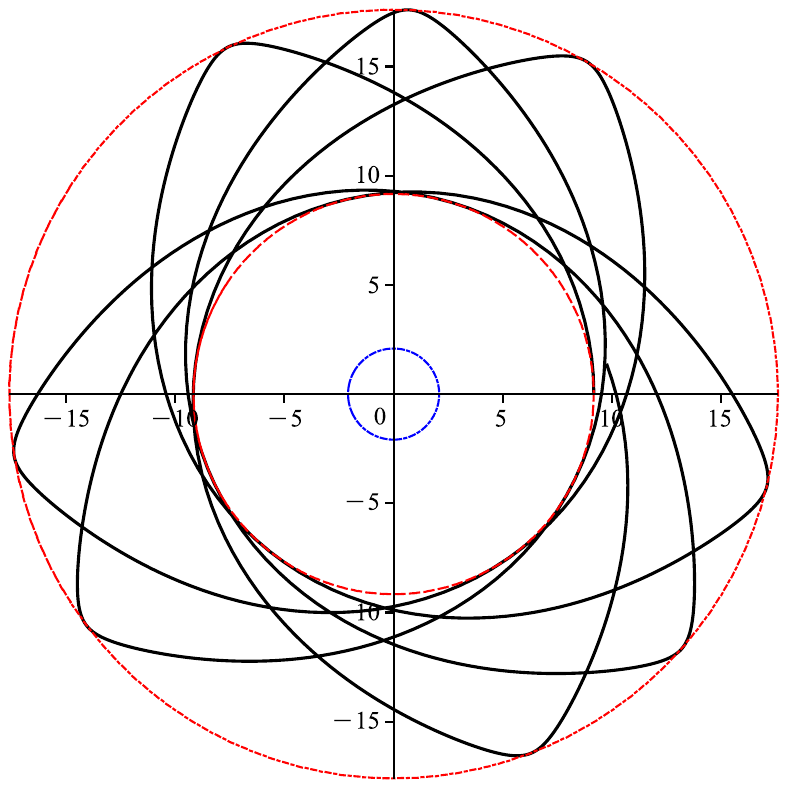}
\caption{{\it Left panel.} Bound orbit of a particle with energy $E^2=1.08$ (the red horizontal line). {\it Right panel.}  Bound orbit of a particle with energy $E^2=1.2$ (the cyan horizontal line).}
\label{bound1}
\end{figure}

\begin{figure}[H]
\centering
\includegraphics[scale=0.4,trim = 2cm 12cm 0cm 1cm]{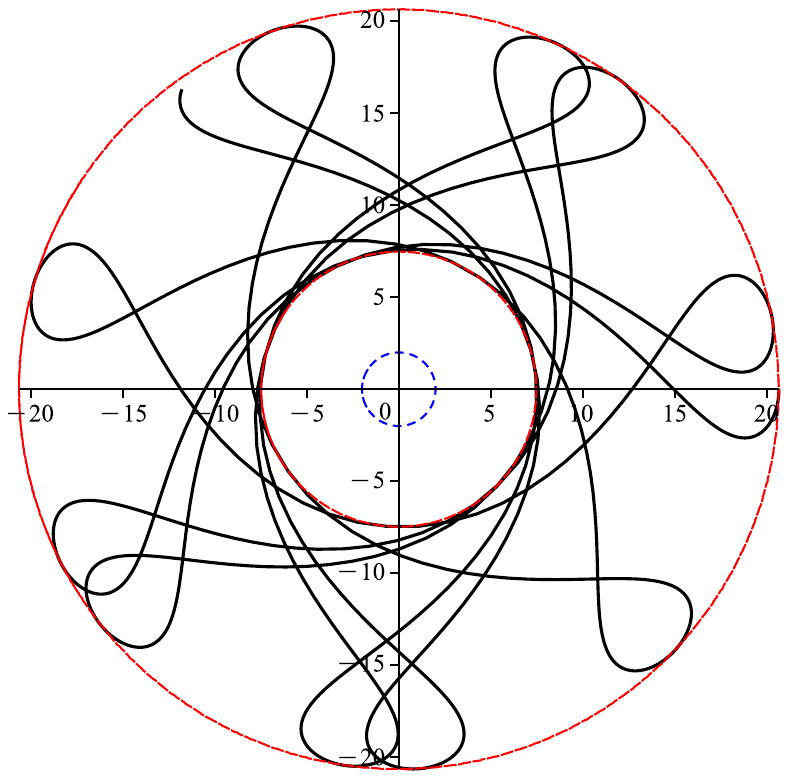}
\includegraphics[scale=0.4,trim = 2cm 12cm 0cm 1cm]{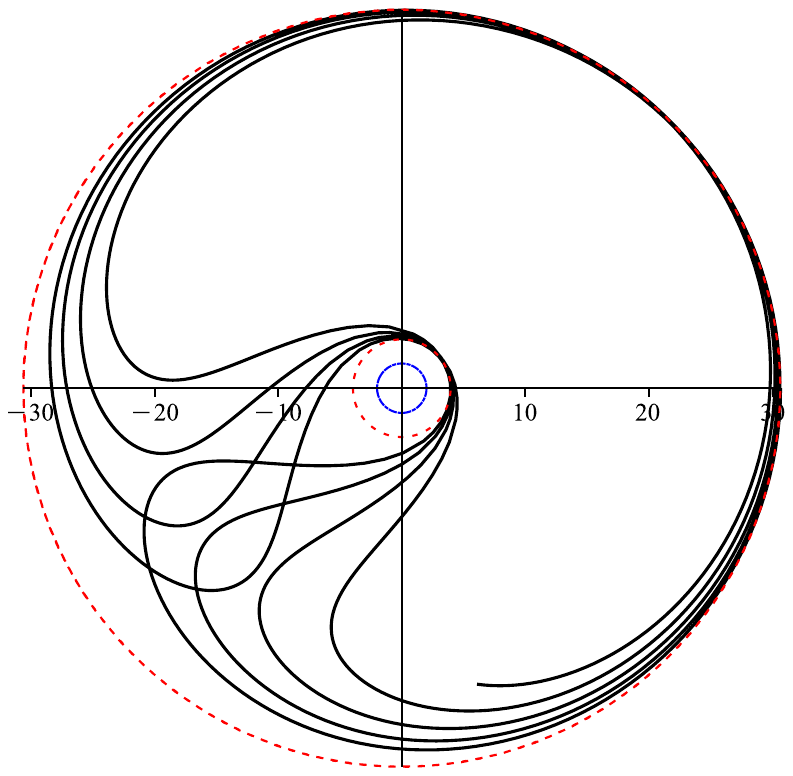}
\caption{{\it Left panel.} Bound orbit of a particle with energy $E^2=1.4$ (the blue horizontal line). {\it Right panel.} Bound orbit of a particle with energy $E^2=2.5$ (the green horizontal line).}
\label{bound2}
\end{figure}

\item{} {\it Capturing orbit.} The particle with $E^2 \leq V_{max1}$ which starts its journey in $r_0$ situated in the blue region and the particle with $V_{max1} \leq E^2 \leq V_{max2}$ with $r_0$ in the green region will be captured by the black hole (see the figure \ref{capture}). 
\item{} {\it Escaping orbit.} The particle with $E^2 \leq V_{max2}$ and $r_0$ is in the red region, will escape toward the cosmological horizon (see the figure \ref{escape}).
\begin{figure}[H]
\centering
\includegraphics[scale=0.4,trim = 2cm 12cm 0cm 1cm]{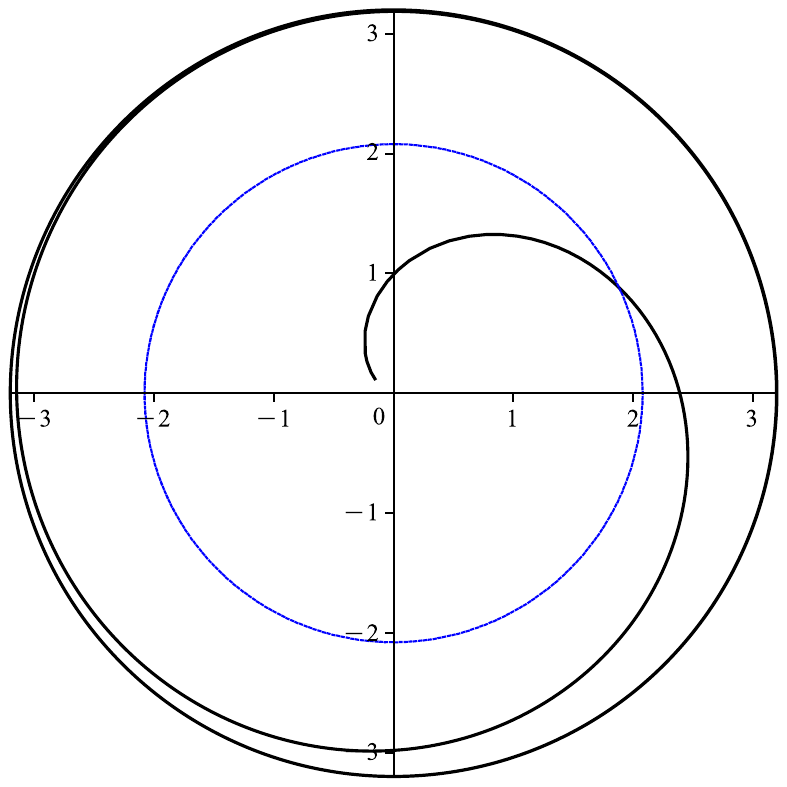}
\includegraphics[scale=0.4,trim = 2cm 12cm 0cm 1cm]{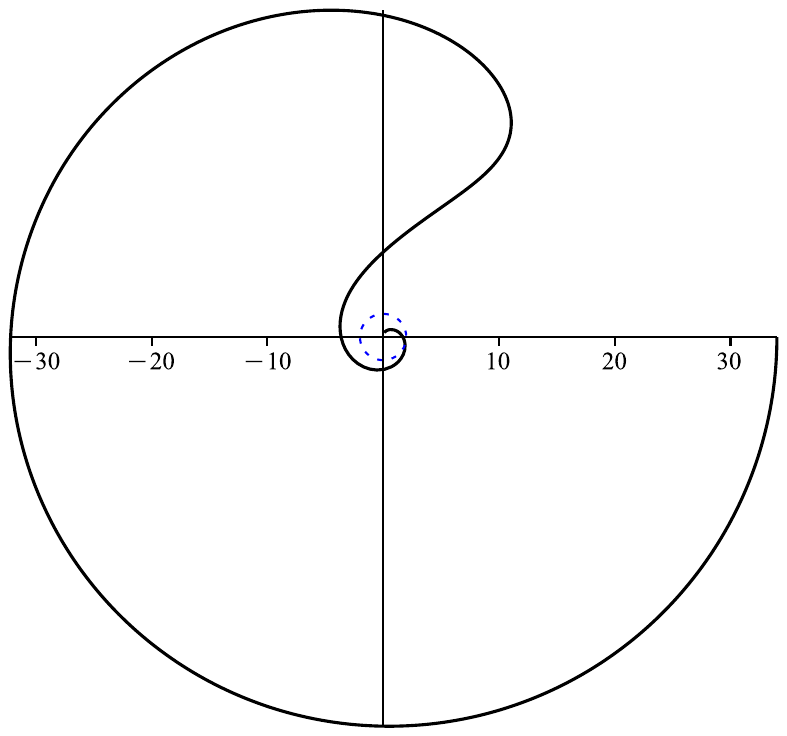}
\caption{{\it Left panel.} First unstable circular orbit of a particle with energy  $E^2= V_{max1} = 2.72$. {\it Right panel.}  Capturing orbit of the particle with energy $E^2=2.9 < V_{max2} = 3.17$ and $r_0$ in the green region.}
\label{capture}
\end{figure}

\begin{figure}[H]
\centering
\includegraphics[scale=0.4,trim = 2cm 12cm 0cm 1cm]{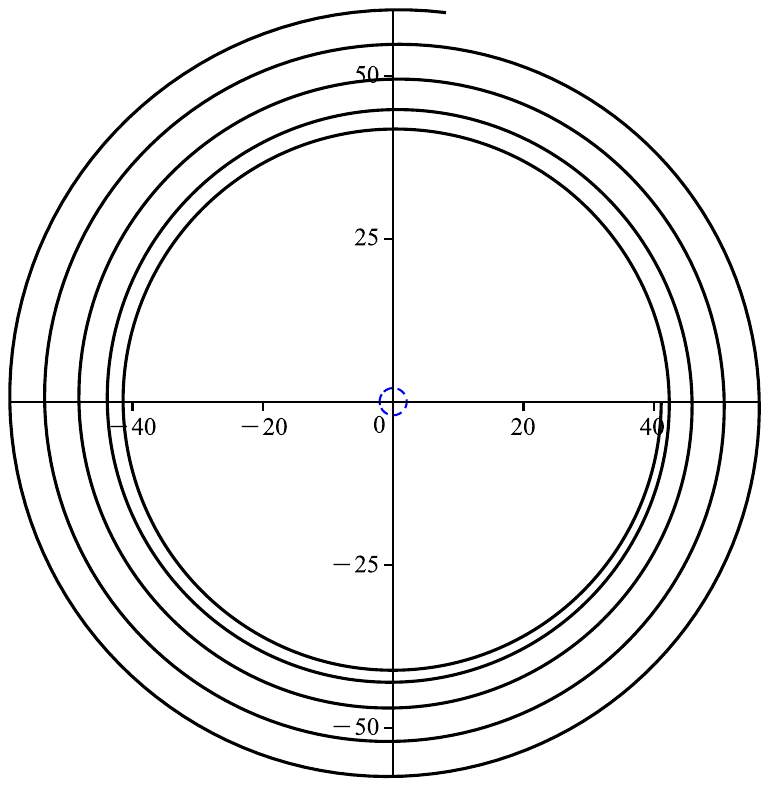}
\caption{Escaping trajectory of the particle with $E^2 = 3 < V_{max2} = 3.17$ and $r_0$ in the red region.}
\label{escape}
\end{figure}
\end{itemize}

The presented results lead to important conclusions on the effect of the model's parameters on the particle's trajectories in the equatorial plane.
As it can be noticed in the figure \ref{PotEq1}, the effective potential is strongly depending on the parameters $B_0$ and $k$. Firstly, using the right panel of the figure \ref{PotEq1}, let us fix the value of $B_0$ and discuss the corresponding orbits for different values of $k \in [0 , 1/(8M)$. The bounded orbits exist for potentials as the one represented in the figure \ref{PotEq2}, namely when the value of $k$ is small and the potential has a minimum value between two maxima. The corresponding bounded trajectories are represented in the figures \ref{bound1} and \ref{bound2}.
As we can notice in the right panel of figure \ref{PotEq1}, there is a critical value of $k$, denoted by $k_*$, above which the second maximum of the potential vanishes and the particle is following either a capturing orbit or an escaping one, depending on the starting point. As an example, with $k=0.04$ and keeping the rest of the numerical values as in the figure \ref{PotEq2}, the particle with the energy $E^2= 1.4$, whose bounded trajectory was given in the figure \ref{bound2}, once $k$ increases to $k=0.04$, has either a capturing orbit or an escaping one.

Secondly, let us discuss the effective potential and the corresponding orbits for different values of $B_0$. For $k=0$, one is recovering the Ernst spacetime which is not asymptotically flat and has one event horizon in $r_h =2M$. Once the parameter $k$ comes into place, the potential vanishes on the cosmological horizon to which particles with suitable energies can escape. Using the left panel of the figure \ref{PotEq1}, where the value of $k$ is fixed, one may noticed that the potential's second maximum, if exists, is strongly increasing with $B_0$. For each value of $k$, one has to find the range of $B_0$ for which there is a potential well in which the particles can be trapped. There is a critical value $B_*$ above which there are no bounded orbits. The value of $B_*$ is depending on $k$.

Finally, let us notice that the magnetic field is responsible for the existence of two types of bounded orbits: the curly ones and
trajectories with no curls. Thus, in cases where $\dot{\varphi}$ defined in (\ref{varphi0}) changes sign in $r_* \in [r_- , r_+ ]$, where
\[
r_* = \sqrt{\frac{L}{(\varepsilon - L B_0) B_0}} \, ,
\]
the corresponding trajectory is curly (see the figure \ref{bound2}).
For $L<0$ and $\varepsilon >0$ or $L>0$ and $\varepsilon <0$, the quantity $\dot{\varphi}$ given in (\ref{varphi0}) is not changing the sign and the corresponding trajectories are without curls.

\section{Charged particles in weakly magnetized Kiselev BH}

\subsection{The magnetic weak-field approximation}

In the weak field limit, one considers magnetic fields that are sufficiently weak not to influence the spacetime curvature. In this case, the metric is basically that of the Kiselev geometry (for $\Lambda\ra 1$), however, the magnetic potential $A_{\varphi}$ retains its expression given in (\ref{A3}).
According to data \cite{Daly}, the magnetic induction was estimated to typically belong in the range $B_G \in [ 10^4 , 10^8]$ (G). For a black hole with the mass $M = 10^{9} M_S$, which is typical for active galactic nuclei, this range of magnetic fields is much smaller than the critical magnetic induction that can influence the space-time geometry \cite{Frolov}
\[
B_G \sim \frac{c^4}{G^{3/2} M_S} \cdot \frac{M_S}{M} \sim 10^{19} M_S/M \sim 10^{10} \; (G)
\]
On the other hand, for charged test particles, the quantity $\varepsilon = q/m$ is very large and therefore the Lorentz force cannot be neglected even for weak magnetic fields. Following \cite{lim}, one can obtain the weak-field approximation in the equations of motion for a charged test particle by taking the limit $B_0\ra 0$ (hence $\Lambda \to 1$) while keeping the quantity $B_0\varepsilon$ finite in the full equations of motion given in Section $3.1$. More specifically, one can take $B_0=\frac{b}{\varepsilon}$, where $b$ is a constant parameter and consider an expansion of the equations of motion (\ref{rEq}) and (\ref{tEq}) in powers of $\frac{1}{\varepsilon}$. Therefore, in the first order of approximation one has $B_0\ra0$, hence $\Lambda\ra 1$ in the final geometry (\ref{ds2}), while the effect of the magnetic field is manifested by means of the Lorentz force in the motion of electrically charged particles.

\begin{figure}[H]
\centering
\includegraphics[width=0.5\textwidth]{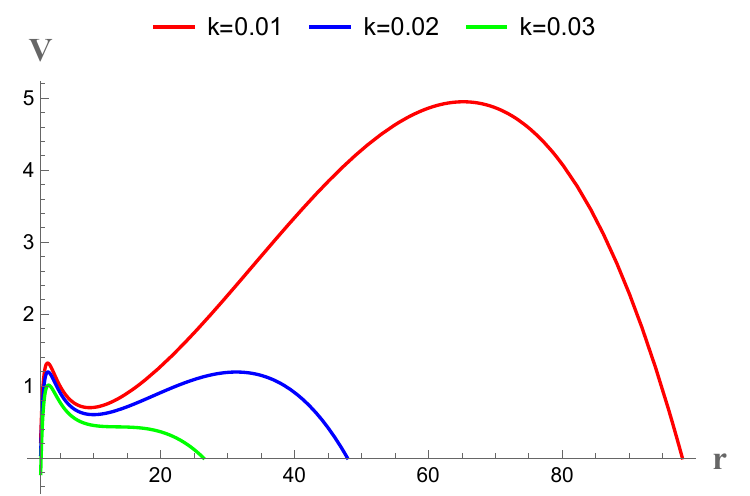} \hspace{0.2cm}
\includegraphics[width=0.45\textwidth]{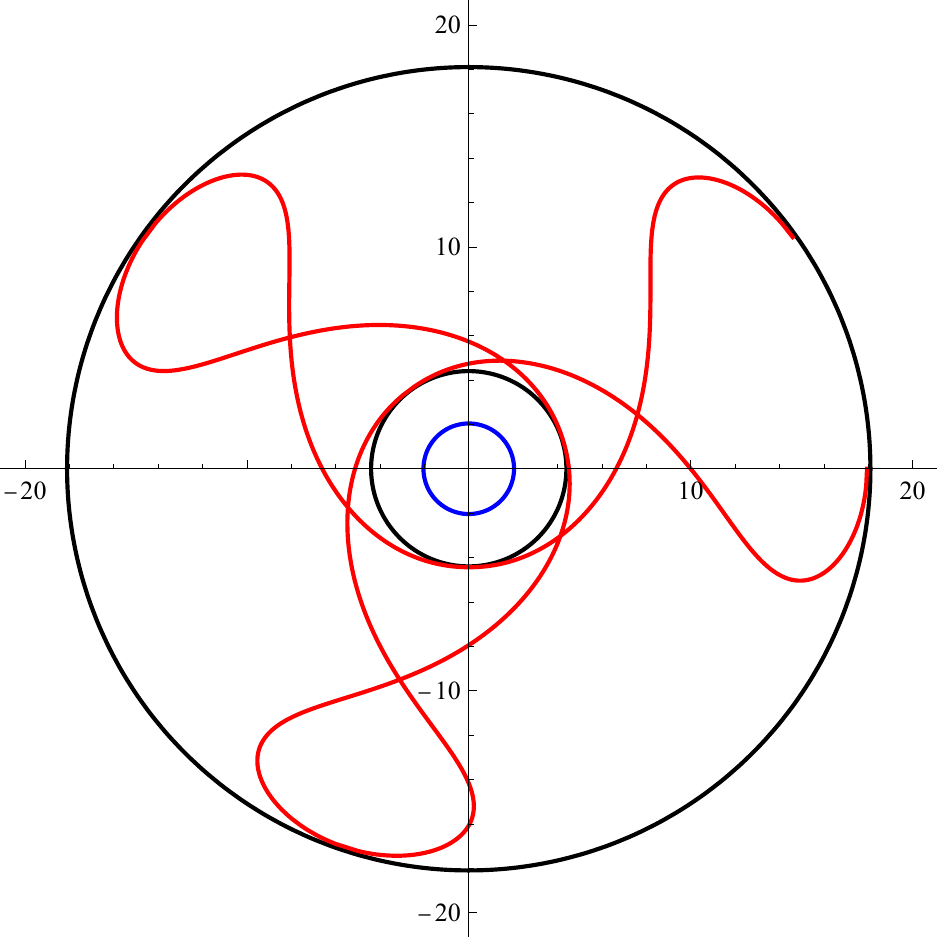} 
\caption{{\it Left panel.} The effective potential (\ref{Vmag0}) for different values of the parameter $k$. {\it Right panel}. A bounded trajectory in the equatorial plane. The numerical values of the parameters are: $M=1$, $k=0.01$,
$B_0 = 0.06$, $L=6$, $\varepsilon =1$, $E^2=1.1$.}
\label{TrajWeak}
\end{figure}

In the equatorial plane $\theta = \pi/2$,
the relations (\ref{phi}), (\ref{rdot}) and (\ref{Vmag}) become
\begin{equation}
\dot{\varphi} = \frac{1}{r^2} \left[ L - \varepsilon B_0r^2 \right] ,
\label{phi0}
\end{equation}
\begin{equation}
\dot{r}^2   = E^2 -V_{mag0} 
\label{rdot0}
\end{equation}
and
\begin{equation}
V_{mag0} =  f(r)  \left[ 1 + \frac{1}{r^2} \left( L - \varepsilon B_0 r^2 \right)^2 \right]
\label{Vmag0}
\end{equation}
with $f(r)$ given in (\ref{kisf}).
The particle's trajectory can be obtained by integrating the relation
\begin{equation}
\frac{dr}{d \varphi} = r^2 \sqrt{E^2 - V_{mag0}} \left[ L - \varepsilon B_0 r^2 \right]^{-1}
\label{traj}
\end{equation}
The potential (\ref{Vmag0}) is represented in the left panel of the fig \ref{TrajWeak}, for different values of the quintessence parameter $k$, and it allows the same type of trajectories as the ones discussed in the subsection 3.3. Thus, one may notice that the effective potential is vanishing on the horizons. When the quintessence parameter $k$ is increasing, the second maximum of the potential disappears and the particle cannot be trapped on a bounded trajectory.
An example of a bounded orbit in the equatorial plane is given in the right panel of the figure \ref{TrajWeak}. The black circles correspond to the turning points $r_1$ and $r_2$, solutions of the equation $E^2 = V_{mag0}$. 

As it can be noticed in the right panel of the figure \ref{TrajWeak}, the presence of
the magnetic field leads to a curly-type bounded orbit since
$\dot{\varphi}$ defined in the equation (\ref{phi0}) changes the sign in $r_* = \sqrt{L/(\varepsilon B_0)}$ and $r_* \in [r_1 , r_2 ]$. For particles with positive $L$ and negative $\varepsilon$, $\dot{\varphi}$ is always positive and the corresponding trajectory has no curls.

\subsection{Perturbed circular orbits in the equatorial plane}

In the equatorial plane ($\theta = \pi/2$), the radial equation (\ref{rEq}),
in the weak field approximation $\Lambda =1$, can be written as
\begin{equation}
\ddot{r} \,  =
\frac{2 f - r f^{\prime}}{2 r^3} \left(  L - \varepsilon B_0 r^2  \right)^2  + \frac{2 \varepsilon B_0 f}{r}  \left(  L - \varepsilon B_0 r^2  \right) - \frac{M}{r^2} + \frac{k}{2}
\label{rEqw}
\end{equation}
One may notice that the Newtonian gravitational force $M/r^2$
directed along the radius toward the black
hole considered in \cite{Frolov} is decreased by a constant quintessence contribution.
In the particular case $\theta = \pi/2$ and $\Lambda = \Lambda_0 =1$, one has the effective potential (\ref{Vmag0}) and the relation
(\ref{phi0}).

The condition for a circular orbit $V^{\prime}_{mag0} (r_0) =0$ leads to the following expression of the angular momentum
\[
L = \left. \frac{r \left[ - \varepsilon B_0 r^2 f^{\prime}  \pm \sqrt{[4 \varepsilon^2 B_0^2 f^2-{f^{\prime}}^{2} ] r^2 +2 r f f^{\prime}} \right]}{2f-r f^{\prime}} \right|_{r=r_0}
\]
and one has to impose that the above expression is a real quantity for $f$ given in (\ref{kisf}). A physical range for the circular orbit radius is $r_0 \in [ 4M , \sqrt{2M/k}]$ and there are no constrains for $B_0$.
Once $r_0 > r_* = \sqrt{2M/k}$, the magnetic induction should exceed a minimum value so that $L$ is real.
However, there is a maximum value of $B_0$ \cite{Esteban} and a maximum value of the quintessence parameter $k$ above which there are no stable circular orbits (see the left panel of the figure \ref{TrajWeak}).

In the followings, we shall apply to ({\ref{rEqw}) the procedure described in \cite{Frolov}, where the Newtonian gravitational force $g =- M/r^2$ is considered as a perturbation for $r \gg 2M$. In our case, $g$ is replaced by $g_k = - M/r^2 + k/2$ and we focus on the region of $r$ where bounded orbits exist and $g_k$ can be treated as a perturbation. For $r_0 < r_* = \sqrt{2M/k}$, the quantity $g_k$ is negative, at $r_0 = r_*$ it vanishes, while for $r_0 > r_*$ it becomes positive.
  
As in \cite{Frolov}, let us consider a point $(r_0 , \varphi_0)$ and introduce the local Cartesian coordinates $(x, y)$ near it.
For $x \ll r_0$ and $y \ll r_0$ one can work in the first order approximation
\begin{equation}
r \approx r_0 + y \; , \quad \varphi \approx \varphi_0 - x 
\label{xya}
\end{equation}
with the radius $r_0$ obtained in the zero order approximation as being
$r_0 = \sqrt{L/(\varepsilon B_0)}$.
To first order in $x$, $y$ and $k$, the equations (\ref{rEqw}) and (\ref{phi0}) lead to the following system of equations
\begin{eqnarray}
&& \ddot{y} +  \Omega^2 y = - \gamma \nonumber \\*
& & \dot{x} = \frac{ 2 \varepsilon B_0}{r_0} y
\label{xySysa}
\end{eqnarray}
where we have introduced the notations
\begin{eqnarray}
& & \gamma =\frac{M}{r_0^2} - \frac{k}{2}  \nonumber \\*
& &
\Omega^2 =  4 \varepsilon^2 B_0^2 \left( 1 - \frac{2M}{r_0}- k r_0 \right) 
\label{gam}
\end{eqnarray}
The solutions of the system (\ref{xySysa}) are given by the expressions
\begin{eqnarray}
& &
y( \tau ) = - \frac{\gamma}{\Omega^2} + \eta \cos ( \Omega \tau ) \nonumber \\*
& &
x ( \tau ) = - \frac{2 \varepsilon B_0 \gamma}{\Omega^2 r_0} \tau +
 \frac{2 \varepsilon B_0 \eta}{\Omega r_0} \sin ( \Omega \tau ) 
\label{xysola}
\end{eqnarray}
where $\eta$ is an integration constant and one has to consider $\eta \ll r_0$ so that the first order approximation is valid. The circular orbit is stable since the frequency $\Omega$ is real.

\subsection{Trochoidlike trajectories in the equatorial plane}

The shape of the perturbed trajectory in the equatorial plane is depending on the metric parameters $M$ and $k$ whose values are affecting the expressions of $\Omega$ and $\gamma$ defined in (\ref{gam}). Thus, the frequency $\Omega$ is decreassing with increasing $M$ and/or $k$. In the important parameter $\gamma$, the terms $M/r_0^2$ and $k/2$ compete against each other. Therefore, $\gamma$ may be either positive or negative and also the ratio $\eta_0 = \gamma/\Omega^2$ which  separates the curly trajectory from the one without curls.
 The line on the $(x, y)$ plane described by the solution (\ref{xysola}) is
called a trochoid and for $\eta = \eta_0$ it turns into a cycloid. 
The presence of $k$ is leading to a critical value of the radial coordinate, $r_* = \sqrt{2M/k}$.
For $r_0 < r_* = \sqrt{2M/k}$, meaning that the gravity is dominant since $M/r_0^2>k/2$, both $\gamma$ and $\eta_0$ are positive quantities. In the opposite case corresponding to $k/2>M^2/r_0^2$, i.e. $r_0 > r_* = \sqrt{2M/k}$, the parameters $\gamma$ and $\eta_0$ are negative and the shape of the trajectory is changed, as it can be noticed in the figure \ref{Cyc}. In the left panel, we give the parametric plot of the trajectory corresponding to positive $\gamma$ and $\eta  = 2 \eta_0 = 2 \gamma/\Omega^2$. For a comparison, the blue line without curls corresponds to the case analyzed in \cite{Frolov}, where $k=0$ so that $\gamma = M/ r_0^2$ and $\Omega \approx 2 \varepsilon B_0$. One may also notice that the particle's orbit is curled toward the black hole and this happens because the attractive term produced by gravity dominates over the repulsive one produced by quintessence.
In the right panel, we consider the opposite case where the quintesence contribution becomes dominant and both $\gamma$ and $\eta = \eta_0$ are negative. The particle's orbit is curled outward the black hole. This case has no equivalence in Ernst spacetime analyzed in \cite{Frolov}. In the particular situation when $r_0 = r_*$ so that $\eta = 0$, we recover the motion in a flat spacetime and the particle has a helicoidal trajectory as the one in the figure \ref{Mel2}.

\begin{figure}[H] 
\centering
\includegraphics[width=0.45\textwidth]{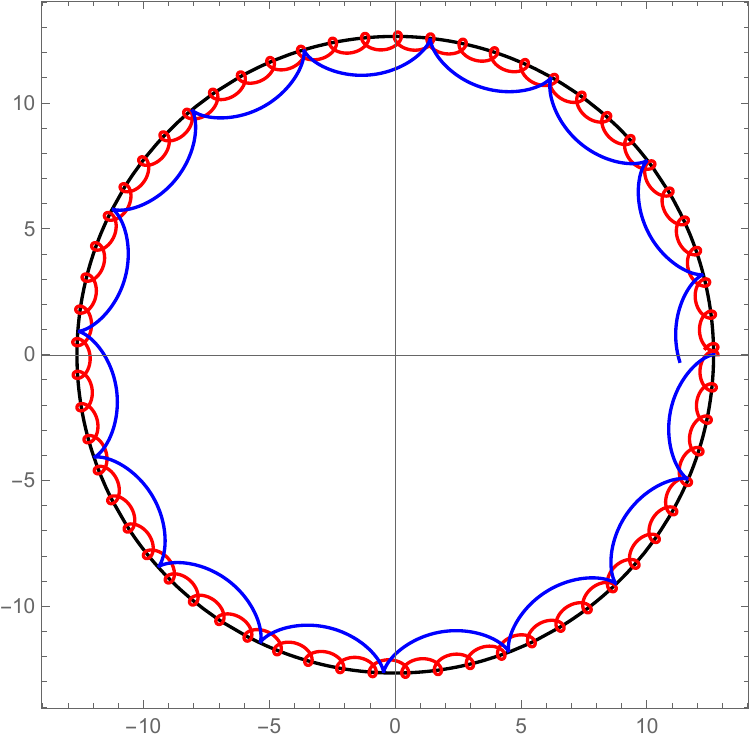}  \hspace{0.2cm}
\includegraphics[width=0.45\textwidth]{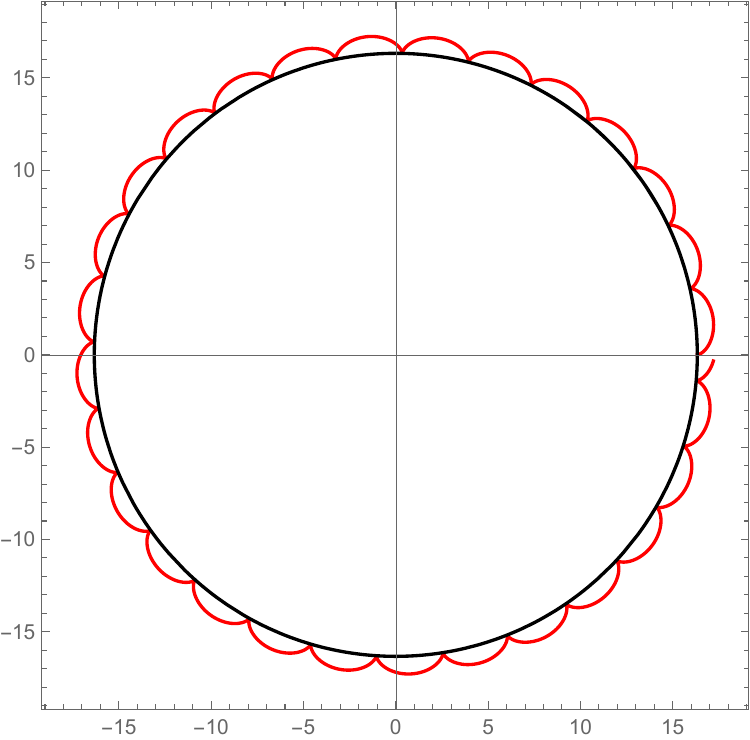}
\caption{{\it Left panel}. The red graphic is the parametric plot of (\ref{xya}) with the solutions (\ref{xysola}) for pozitive $\gamma$ defined in (\ref{gam}) and $\eta = 2 \eta_0 = 2 \gamma / \Omega^2$. 
The numerical values of the parameters are: $M=1$, $k=0.01$, $L=8$, $\varepsilon =1$, $B_0=0.05$. The blue curve is for $k=0$ and $\eta = \eta_0$.
{\it Right panel}. Parametric plot of (\ref{xya}) with the solutions (\ref{xysola}) for $\gamma<0$. 
The numerical values of the parameters are: $M=1$, $k=0.01$, $L=8$, $\varepsilon =1$, $B_0=0.03$, $\eta = \eta_0 <0$. The black circle represents the unperturbed circular orbit of radius $r_0=\sqrt{L / (\varepsilon B_0)}$.}.
\label{Cyc}
\end{figure}

\subsection{The three-dimensional approach. Mathieu functions and stability charts}

Secondly, let us analyze the more involved case of the general system of equations (\ref{rEq}), (\ref{tEq}) and (\ref{phi}), in the weak magnetic field approximation. For $\Lambda =1$, these equations get the expressions:
\begin{eqnarray}
&&
\ddot{r} \,  = \, \frac{2 f -r f^{\prime}}{2} r \dot{\theta}^2 +
\frac{2 f -r f^{\prime}}{2 r^3 \sin^2 \theta}
\left[ L - \varepsilon B_0 r^2 \sin^2 \theta\right]^2 + \frac{2 \varepsilon B_0 f}{r} \left[ L - \varepsilon B_0r^2 \sin^2 \theta \right] - \frac{f^{\prime}}{2}
\nonumber \\*
& &
\ddot{\theta} \,  = \, 
- \frac{2}{r} \dot{r} \dot{\theta} + \frac{\cot \theta}{r^4 \sin^2 \theta } 
 \left[ L - \varepsilon B_0r^2 \sin^2 \theta \right]^2 + \frac{2 \varepsilon B_0 \cot \theta}{r^2}   \left[ L - \varepsilon B_0r^2 \sin^2 \theta \right]
\nonumber \\*
& & \dot{\varphi} = \frac{1}{r^2 \sin^2 \theta} \left[ L -\varepsilon B_0r^2 \sin^2 \theta\right]
\label{weak} 
\end{eqnarray}
and the effective potential reads
\begin{equation}
V =  f(r) \left[ 1 + \frac{\left( L - \varepsilon B_0 r^2 \sin^2 \theta \right)^2}{r^2 \sin^2 \theta} \right]
\end{equation}
In the first order approximation, one can consider
\begin{equation}
r \approx r_0 + y \; , \quad \theta \approx \frac{\pi}{2} - z \; , \quad \varphi \approx \varphi_0 - x 
\end{equation}
where $r_0 = \sqrt{L/(\varepsilon B_0)}$, in the zero order approximation.
To first order in $x$, $y$ and $z$, the equations (\ref{weak}) lead to the system of equations (\ref{xySysa}) with the solutions (\ref{xysola}) and to the $z-$equation
\begin{equation}
\ddot{z} =  \frac{ 4 \varepsilon^2 B_0^2}{r_0} z y 
\end{equation}
i.e.
\begin{equation}
\ddot{z} + \frac{ 4 \varepsilon^2 B_0^2}{r_0}
\left[ \eta_0 - \eta \cos ( \Omega \tau )  \right] z = 0
\label{Mat}
\end{equation}
where $\eta_0 =  \frac{\gamma}{\Omega^2}$. This is a Mathieu's type equation whose general form is \cite{Mathieu}
\begin{equation}
\ddot{z} + \left[ a - 2 q \cos ( 2 t )  \right] z = 0
\label{mathieu}
\end{equation}
the solutions being the Mathieu characteristic functions denoted as $ce_n(q, t)$ and $se_n(q, t)$, where $n$ is an integer. These are periodic functions, with the period $\pi$ or $2\pi$ and exist only for specific values of parameters $a$ and $q$.

In our case, the solutions of (\ref{Mat}) are the Mathieu functions:
\begin{equation}
z( \tau ) = C_1 MathieuC \left[ a , q,  \frac{\Omega \tau}{2} \right] + C_2 Mathieu S \left[ a , q ,  \frac{\Omega \tau}{2} \right]
\label{zsola}
\end{equation}
with 
\begin{equation}
a =\frac{16 \varepsilon^2 B_0^2 \eta_0}{r_0 \Omega^2} \; , \quad
q = \frac{8 \varepsilon^2 B_0^2 \eta}{r_0 \Omega^2} .
\end{equation}
The constants $a$ and $q$ are usually referred as characteristic
number and parameter, respectively.
From the general theory of the Mathieu's functions \cite{Mathieu}, we know that these are of the form
\[
u \sim e^{i r x} g (x) \, ,
\]
where $g (x)$ is a periodic function. Depending on the parameters $a$ and $q$, the so-called Mathieu Characteristic Exponent $r$ may be real or imaginary.
When $r$ becomes imaginary, the real and imaginary parts of the Mathieu functions are exponentially growing as a result of a parametric resonance.
Even though there exists a comprehensive literature about the Mathieu equation and its solutions, the crucial problem is
to find conditions for the parameters such that the solutions remain bounded. The stability charts give the transition curves in the parameters plane which separate the regions corresponding to a stable motion from those of instability.

\begin{figure}[H]
\centering
\includegraphics[scale=0.4,trim = 2cm 12cm 2cm 2cm]{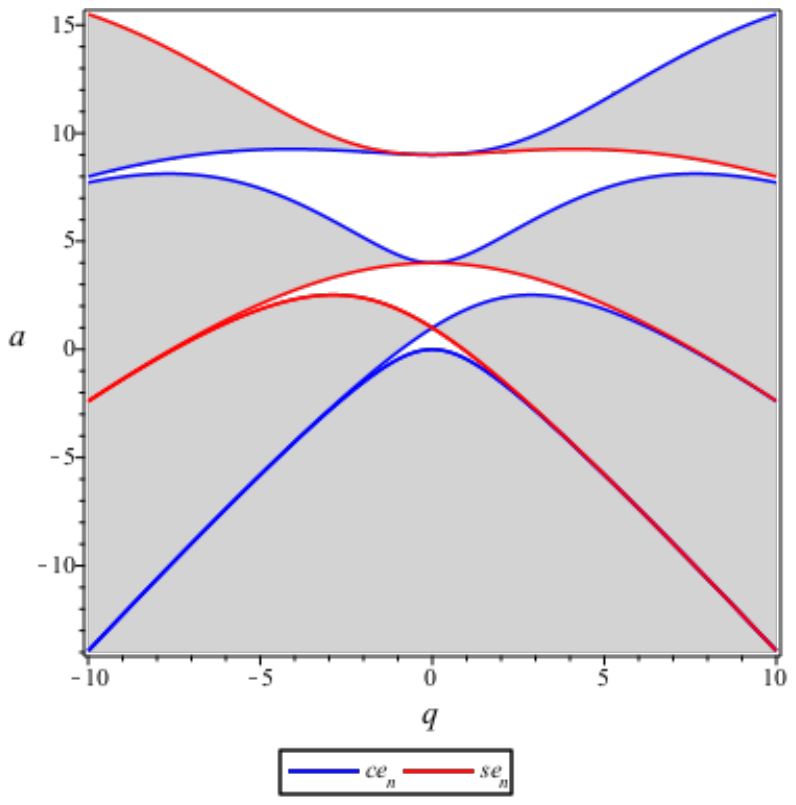} \hspace{0.2cm}
\includegraphics[scale=0.4,trim = 2cm 12cm 2cm 2cm]{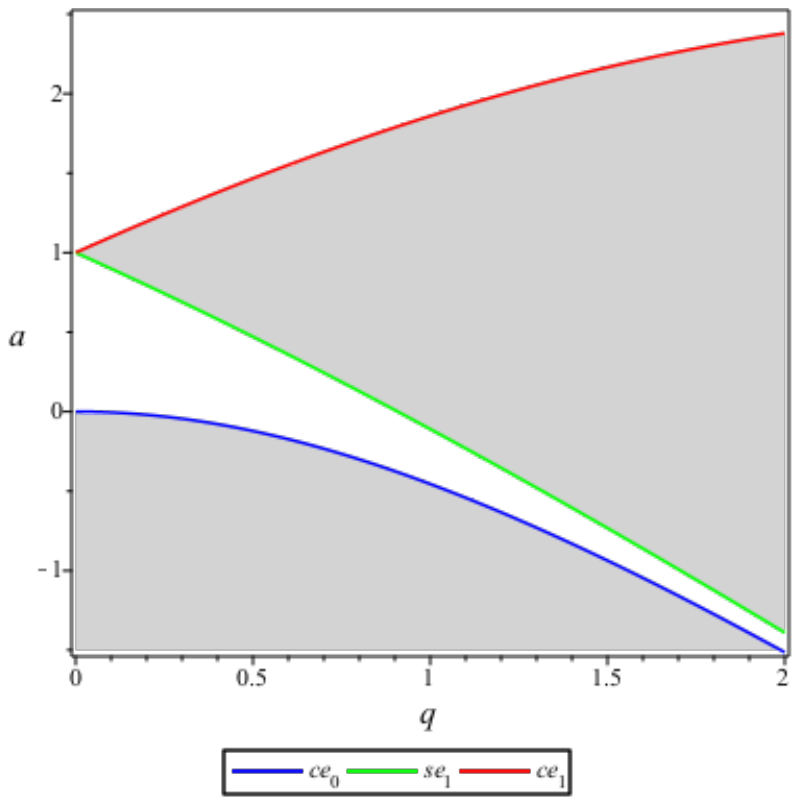}
\caption{Stability chart for the Mathieu equation (\ref{mathieu}). Stability regions are the white areas, while the grey ones correspond to instability regions.}
\label{stability}
\end{figure}

In our case, for a given value of $a$, using the diagrams as the ones represented in the figure \ref{stability}, we find the range of $\eta$ so that the corresponding $q$ is in a stability region. In the figure \ref{Cyc3D}, in the left panel, we give the 3-dimensional parametric plot corresponding to the solutions (\ref{xysola}, \ref{zsola}) and the projection on the $xy$ plane is given in the right panel.
For the chosen numerical values, one has $a = 0.077$ and $q = 0.15$ and therefore we are in the first stability region.  

Finally, one has to point out that the Mathieu parameters $a$ and $q$ which play an important role in the stability of particle's trajectory are depending on the values of the metric parameters $M$ and $k$. Thus, if $r_0 < r_* = \sqrt{2M/k}$, i.e. $M/r_0^2>k/2$, the parameter $a$ in the Mathieu functions is positive, since $\eta_0 = \gamma$ defined in (\ref{gam}) is positive. Thus, one has large regions of stability, as it can be noticed in the figure \ref{stability}. In the presence of quintessential matter, if $k/2>M/r_0^2$ and $\eta_0$ becomes negative leading to a negative $a$, there are very narrow ranges for the parameter $q$ where the solutions are stable and bounded (see the figure \ref{stability}). In this case, the Mathieu Characteristic Exponent becomes imaginary and the Mathieu functions are exponentially growing. Thus, one may conclude by saying that the presence of quintessence is strongly affecting the particle's trajectory which becomes unstable.

\begin{figure}[H]
\centering
\includegraphics[width=0.4\textwidth]{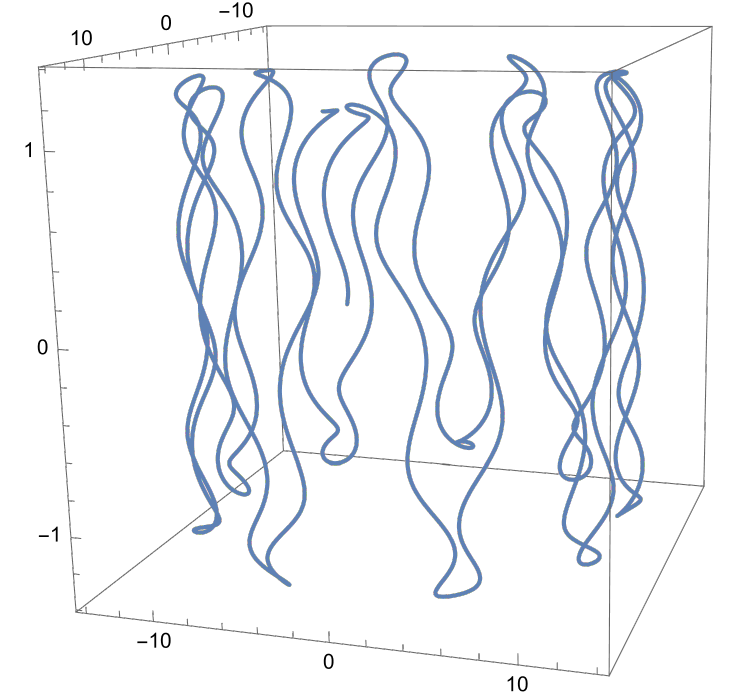} \hspace{0.2cm}
\includegraphics[width=0.4\textwidth]{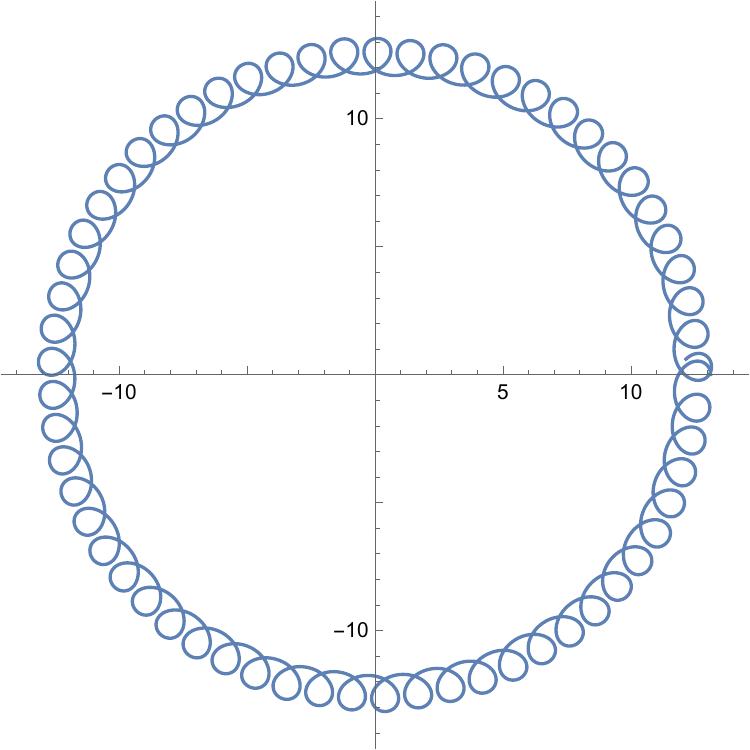}  
\caption{{\it Left panel.} Three dimensional parametric plot with the solutions (\ref{xysola}, \ref{zsola}) . {\it Right panel}. The projection on the $xy-$plane. The numerical values of the parameters are: $M=1$, $k=0.01$, $L=8$, $\varepsilon =1$, $B_0=0.05$, $\eta = 0.7 > \eta_0$}
\label{Cyc3D}
\end{figure}

\section{Conclusions}

The present paper deals with charged particles motion around a Schwarzschild black hole immersed in an axisymmetric
magnetic field, in the presence of quintessential matter. A comparison with the charged particles trajectories in the Melvin magnetic universe \cite{lim} is pointing out the existence, in both cases, of curly trajectories which is characteristic for the charged particles experiencing an inward gravitational force which is counteracted by an outwards Lorentz force.

Starting with the general Lagrangian (\ref{Lag}), we derive the field equations and the effective potential whose plot gives us important information on the allowed regions of motion and equilibrium points.
The field equations are not exactly solvable and we have used numerical methods to represent the trajectories. In particular cases, the explicit form of geodesic motion can be obtained in terms of elliptic integrals \cite{bat}.
In our case, the equations (\ref{rEq}), (\ref{tEq}), (\ref{phi}) were solved numerically by using the Maple software and implementing a Runge-Kutta algorithm of 4th order. As a consistency check we used the relation (\ref{rdot})\footnote{Some useful Mathematica notebooks for analyzing the motion of charged particles in various magnetic field configurations can be found at the following repository https://github.com/XyhwX/particle. We thank Martin Kolos for drawing our attention to it.}.
The trajectories both in three dimensions and in the equatorial plane are classified and discussed in detail, pointing out the effect of the model's parameters on their shape.

A special attention is given to the case of a weak magnetic field which can be analytically treated. Firstly, we have considered the orbits in the equatorial plane. The gravitational force $M/r^2$
directed toward the black
hole is decreased by a constant quintessence contribution and bounded trajectories exist only for low values of the quintessence parameter. For circular orbits, the perturbative approach 
reveals a cycloidlike or trochoidlike motion, similar
to those found by Frolov \cite{Frolov} and Lim \cite{lim} in Ernst spacetime. However, one may notice in the figures \ref{Cyc} that the shape of the perturbed circular trajectory is dictated by the parameter $\gamma$ whose sign reflects the relation between the attractive gravitational force and the repulsive contribution induced by the presence of quintessential matter.
The quintessence parameter value is affecting the frequency of the circular motion as well as the transition from a curly to a non-curly trajectory.

Secondly, we have turned to a three-dimensional investigation and worked out, in the first-order approximation, the general system of equations (\Ref{weak}).
The radial equation reduces to that of a harmonic oscillator driven by a periodic force, while the perturbed motion in the $\theta$ direction is governed by a Mathieu-type equation.
For $r_0 <r_* = \sqrt{2M/k}$, one may find large ranges for the model's parameters for which the Mathieu's functions are bounded and stable. This means that the corresponding orbit of radius $r_0$ is indeed stable even when the particles are perturbed slightly away
from the equator (see the figure \ref{Cyc3D}).
On the other hand, when the quintessence contribution becomes dominant and $kr_0 > 2M/r_0$, 
there are very narrow stability ranges for the Mathieu's parameters. Outside these stability regions, the Mathieu's Characteristic Exponent is complex and the corresponding Mathieu's functions are exponentially increasing.
This means that even though the orbit of radius $r_0$ corresponds to a minimum value of the potential, the trajectory is unstable when the particle is slightly perturbed away
from the equatorial plane.

Our results agree with the ones derived in \cite{Kolos, Kol},
where the small radial and latitudinal oscillations around a stable equatorial orbit have been discussed in detail. Also, the case of the movement confined in the equatorial plane has been investigated in \cite{Frolov}.
The main conclusion is that even though there is a clear resemblance to the Larmor precession in a pure magnetic field, the effect of a weak external uniform magnetic
field combined with a gravitational or/and a quintessence force may be substantial.

\section*{Acknowledgements}
The authors would like to thank the anonymous Referees whose remarks and suggestions helped improve this manuscript.

\end{document}